\documentclass[elsart12,epsf,eqsecnum,graphics,cite]{revtex4}
\def\beq{\begin{equation}}

\def\eeq{\end{equation}}

\usepackage{epsfig,epsf}
\usepackage{graphicx}
\usepackage{grffile}

\begin{document}

%\hoffset-1cm

% Yale printer values
\voffset1.5cm

%\draft{BI-TP 2005/19}
%\preprint{BI-TP 2005/19, CERN-PH-TH-2005/99}
\title{On Angular Correlations and High Energy Evolution.}
\author{ Alex Kovner$^{1}$ and Michael Lublinsky$^{2,1}$}

\affiliation{
$^1$ Physics Department, University of Connecticut, 2152 Hillside
Road, Storrs, CT 06269-3046, USA\\
$^2$Physics Department, Ben-Gurion University of the Negev, Beer Sheva 84105, Israel\\}
%\date{\today}

\begin{abstract}
We address the question to what extent JIMWLK evolution is capable of taking into account angular correlations in a high energy hadronic wave function. Our conclusion is that angular (and indeed other) correlations in the wave function cannot be reliably calculated without taking into account Pomeron loops in the evolution. As an example we study numerically
 the energy evolution of angular correlations between dipole scattering amplitudes in the framework of the large $N_c$ approximation to JIMWLK evolution (the "`projectile dipole model"').
Target correlations are introduced via averaging over (isotropic) ensemble of  anisotropic initial conditions. We find that correlations disappear very quickly with rapidity even inside the saturation radius. This is in accordance with our physical picture of JIMWLK evolution. The actual correlations inside the saturation radius in the target QCD wave function, on the other hand should remain sizable at any rapidity.
\end{abstract}
\maketitle
%%%%%%%%%%%%%%%%%%%%%%%%%%%%%%%%%%%%%%%%%%%%%%%%%%%%%%%%%%
\section{Introduction and Conclusions}
The CMS observation of angular and long range rapidity correlations in the hadron spectrum, the so called "ridge" in proton-proton collisions\cite{cms},  has triggered a lot of discussions in recent literature \cite{recent},\cite{ddgjlv},\cite{KLc},\cite{LR}. 
%A similar if more pronounced correlated structure was observed in gold-gold collisions at RHIC \cite{rhicridge}. There is a variety of candidate explanations for the RHIC observation \cite{rhicexplane}, \cite{glv} most of them utilizing strong radial flow as a collimating mechanism. Although flow measurements have not been reported for the LHC data, it is difficult to imagine that flow will have a significant effect in p-p collisions. Thus the viable explanation should probably not appeal to any collective behavior of produced particles.
In particular the approaches of the three last papers \cite{ddgjlv},\cite{KLc},\cite{LR} are based on the idea that correlated gluon emission is due to the correlations in impact parameter plane preexisting in the incoming wave function of the target and projectile hadrons. Such correlations certainly exist in a hadron wave function, and in the context of high energy evolution can be encoded in the initial conditions for the evolution. It was also argued in \cite{KLc} that the correlations are leading effect in $1/N_c$. The purpose of the present note is to address the question what is the fate of such correlations as the hadron is evolved to high energy. In particular we ask whether these correlations can be studied by evolving the target/projectile wave functions with JIMWLK evolution\cite{jimwlk} (or the Balitsky-Kovchegov (BK) equation, which is its large $N_c$ limit\cite{bk}). To this end we perform simple numerical calculations and supplement them with qualitative analysis based on physics of the JIMWLK evolution. Our numerics is performed in the framework of the dipole model approximation to JIMWLK evolution, and thus is sensitive only to the leading $N_c$ part of JIMWLK, but we believe that, with minor modifications our conclusions are valid for full JIMWLK evolution as well. 

Our conclusions are the following. We find that the JIMWLK evolution leads to exponentially quick disappearance of correlations (including angular correlations relevant for gluon emission) with rapidity. This disappearance is straightforward to understand. It is the consequence of the fact that JIMWLK evolution is valid only for color field modes with transverse momenta smaller than the saturation momentum where one indeed {\it does not expect} correlations to be present. As was discussed in \cite{KLc}, we expect correlations to be present only for points in transverse plane within the saturation radius of each other, and therefore for momentum modes greater or equal to the saturation momentum. The evolution of these modes even in a dense hadronic wave function is not governed by the JIMWLK evolution but rather by KLWMIJ evolution\cite{klwmij} for $k\gg Q_s$ and by the Reggeon Field Theory (RFT) including Pomeron loops\cite{pomloops} for $k\sim Q_s$.
 
The failure of JIMWLK to properly account for correlations should be understood in the following way.
The evolved wave function of a hadron does indeed contain correlations in impact parameter space, even in the leading order in $1/N_c$ expansion. This has been shown analytically and numerically for angular independent correlations within Mueller's dipole model\cite{dipole} in \cite{yoshi}, and there is every reason to expect that this is also the case for angle dependent correlations of interest to us. However JIMWLK evolution approximates the  scattering amplitude of two dipoles on a hadronic target by contributions where the two dipoles scatter on gluons with vastly different rapidities (this has been dubbed "`long range multiple scatterings"' in \cite{foam}). Since the correlations in the wave functions are between gluons which are close in rapidity to each other, these correlations are simply not included in the JIMWLK equation. While JIMWLK approximation does properly account for leading contributions to the scattering amplitude in the saturated regime, it significantly underestimates those for small (smaller than saturation radius) dipoles. To account properly for rapidity evolution of the scattering amplitude of small dipoles, one needs to evolve it with KLWMIJ evolution equation. The KLWMIJ evolution does indeed include the "`short range multiple scattering contribution"\cite{foam} and should correctly describe correlations in impact parameter plane. The obvious complication here is, that since we are interested in a dense target (corresponding to high multiplicity events in the CMS data), it certainly contains low momentum modes which evolve according to JIMWLK. Thus we are faced with the situation where proper treatment of angular correlations at high energy requires us to include both KLWMIJ and JIMWLK evolutions within the same framework, and in this sense we have to deal explicitly with the Pomeron loop effects. In fact this necessity is even more acute, since the Pomeron loops give leading contributions to the evolution of the modes at $k\sim Q_s$, and it is presumably these modes that contribute most to angular correlations.

We thus conclude that future attempts to properly numerically estimate the size of correlations at high energy will require explicit inclusion of the Pomeron loop effects. We note that the notion that it is the Pomeron loops that are crucial for correlations at high energy is not new \cite{private}, here we merely recast this argument in the framework of JIMWLK/KLWMIJ evolution.

The structure of this paper is the following. In Sec. 2 we recap the arguments of \cite{KLc} about angular correlations in gluon emission, recasting them in a somewhat more transparent semiclassical form. In Sec. 3 we present results of our numerical calculations of the evolution of angular correlations in the dipole model (leading $N_c$ JIMWLK). Finally in Sec. 4 we discuss the interpretation of these results based on the physical picture of JIMWLK/KLWMIJ evolution and flesh out the arguments for necessity of Pomeron loops.
 
\section{Angular Correlations}

In our previous  note \cite{KLc} we discussed a simple picture of long range rapidity correlations and angular correlations between particles produced in a collision of two high energy dense objects. This quialitative picture also underlies the calculations of \cite{ddgjlv},\cite{LR}. 
Long range rapidity correlation is an almost trivial consequence of boost invariance of a projectile wave function at high energy.
Consider high energy scattering of a hadronic projectile on a stationary target in the lab frame. In the lab frame, the incoming particles are  very energetic and they scatter by a very small angle with $p^+\gg p_T$. Thus recoil is negligible and eikonal approximation is applicable at high enough energy. 
Since the projectile is very energetic, its wave function is approximately boost invariant.
% The boost invariance is of course only approximate, since at too high energy the rapidity evolution is important, and that introduces rapidity dependence inside the wave function. However for rapidity intervals $\Delta Y<{1\over\alpha_s}$ the evolution is not important\cite{glv}, and thus can be neglected if the produced particles are separated by rapidity interval which is not parametrically large.
In a boost invariant wave function gluon distribution at rapidity $Y_1$ and $Y_2$ are the same. These gluons scatter on exactly the same target, and thus whatever happens at $Y_1$ also happens at $Y_2$. If for a particular target field configuration a gluon is likely to be produced at $Y_1$ at some impact parameter, a gluon is also likely to be produced at $Y_2$ at the same impact parameter, thus leading to  long range rapidity correlations.  
Thus the long range rapidity correlations come practically for free whenever the energy is high enough so that the wave function of the incoming hadron is approximately boost invariant, and there is very little in the actual dynamics of the collision that can affect this feature.  

To understand why angular correlations also naturally arise in the context of high energy let us briefly recap our understanding of the transverse structure of the hadron in the saturation regime. It is convenient to think of the distribution of the (color) electric field configurations in the target. 

The target wave function is characterized by the saturation momentum $Q_s$. The saturation momentum plays a dual role in the hadronic wave function. First, it measures the typical magnitude of electric field in the wave function. The scattering amplitude of a dipole on the target is given in terms of simple parton scattering amplitude $S(x) = Pe^{ig\int dx^+A^-(x)}$ as $N(r) = 1 -\frac{1}{N_c} Tr[S^\dagger(0)S(r)]$. The vector potential is simply related to the electric field as $\partial_iA^-=F^{-i}$. Let us for convenience define electric field integrated over the longitudinal extent of the target, $E_i=\int dx^+F^{-i}$. The dipole scattering amplitude is then given in terms of $gE$, and assuming for illustrative purposes that odd powers of $E$ average to zero in the hadronic ensemble, we have roughly
\begin{equation}\label{amplitude}
N(\vec r)\sim 1-e^{-\frac{1}{2} (g\vec r\cdot \vec E)^2}
\end{equation}
This is of order unity for $r^2_s=Q^{-2}_s=(gE)^{-2}$.

On the other hand it is known that the field components with transverse momenta $p_T<Q_s$ are suppressed in the wave function\cite{satdist}. This means that the electric fields in the target are correlated on the length scale $\lambda\sim Q_s^{-1}$. Thus the saturation momentum doubles up as the inverse of the correlation length of target color fields. Typical field configurations in the target can thus be thought of having a domain like structure of Fig. 1.  
\begin{figure}[ht]
\includegraphics[width=6.cm]{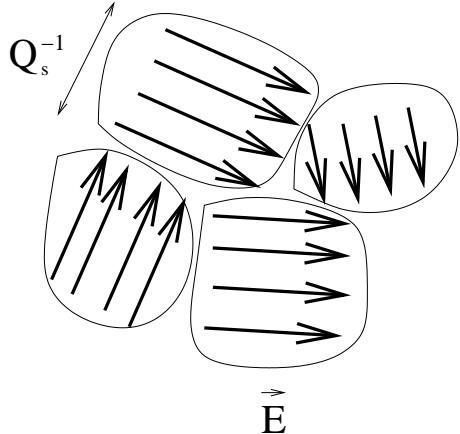} 
\caption{Typical color electric field configuration in the target.}
\end{figure}

Now consider a projectile parton with charge $q$ impinging on one of the domains of the target. While traversing the target field, the parton acquires transverse momentum
\begin{equation}
\delta \vec P= gq\int dx^+ \vec F^{-} = gq\vec E\,.
\end{equation}
A parton at a different rapidity but with the same charge will pick up exactly the same transverse momentum if it scatters on the same "domain". This of course results in positive angular correlation of produced gluons.

We note in passing that this simple picture also explains the fact noted in \cite{KLc} that angular correlations at angle $\phi$ and $\phi+\pi$ have equal strength. At high energy particle production is dominated by gluons. Gluons of course belong to real representation of the gauge group, thus it is equally probable to find an incoming gluon with charge $q$ and charge $-q$ in the projectile wave function at any rapidity. Suppose, for example that a given configuration the color field in the target is in the third direction in color space $E^a_i=E_i\delta^{a3}$, while the gluons in the incoming projectile  corresponds to the vector potential in the second direction $A^2_i$. One can always write $A^2=-i/2(A-A^*)$, where $A=A_1+iA_2$ is positively charged with respect to color charge in the third direction, and $A^-$ is negatively charged. Thus necessarily equal number of gluons in the incoming projectile have opposite sign charges and are kicked in opposite directions while scattering on the target. This produces equal strength correlations at angles zero and $\pi$.
This is not the case for quarks which carry fundamental charges, and it is quite clear that taking into account the projectile quarks will lead to stronger positive angular correlation than the negative one.

Going beyond the qualitative picture described above, the two gluon inclusive production probability discussed in \cite{KLc} is given  by\cite{baier} (see also \cite{multigluons})\footnote{
We will not be interested here in subleading terms discussed in \cite{KLc} which correspond to emission from a single Pomeron, and lead to strong back to back correlations. We thus do not write down these terms in the following.}

%\beq
%\left[{d^2N\over d^2p\,d{\eta}\,d^2k\,d\xi }\,-\,{dN\over d^2k\,d\xi}\,{dN\over d^2p\,d\eta}\right]/{dN\over d^2k\,d\xi}\,{dN\over d^2p\,d\eta}
%\eeq
\begin{equation}
{dN\over d^2pd^2kd\eta d\xi}=\langle\, \sigma(k)\,\sigma(p)\,\rangle_{P,T}  \label{product} 
\end{equation}
with
\begin{equation}
\sigma(k)=\int_{z,\bar z}\,e^{ik(z-\bar z)}\,\int_{x_1, x_2,\bar x_1,\bar x_2} \vec{f}( \bar z - \bar x_1)\cdot \vec{f}( x_1- z)\, 
                 \ \tilde \rho(x_1)[S^\dagger( x_1)-S^\dagger( z)]
                       [S(\bar x_1)-S(\bar z)]\ \tilde \rho(\bar x_1)        \,.  \label{sigma}                              
                                     \end{equation}
Here 
\begin{equation} f_i(x-y)={(x-y)_i\over (x-y)^2}\end{equation}
and  $\tilde\rho\equiv -i T^a\rho^a$. 
In these formulae $\rho^a(x)$ is the valence color charge density in the projectile wave function, while $S^{ab}(x)$ is the eikonal scattering matrix determined by the target color fields. The charge density is normalized such that for a single gluon $\rho^a=gT^a$.
The two gluons here are produced independently of each other, but from exactly the same configuration of color charge sources while scattering on the same target field configuration.

The average in eq.(\ref{product}) denotes averaging over the projectile and the target wave functions. 
The averaging over $\rho$ is understood as averaging over a classical ensemble with a probability distribution function $W_P[\rho]$ \cite{jimwlk}
\begin{equation}\langle O\rangle _P\,=\,\int D\rho \,W_P[\rho] \,O\ .\label{average}\end{equation}
and similarly for the target average.  After averaging over all target and projectile configurations, the single gluon emission amplitude $\langle\sigma\rangle_{P,T}$ must be isotropic. However  for any given configuration it is not isotropic and peaked in one particular direction. This anisotropy produces angular correlation among the emitted gluons as discussed above.
To reiterate, for a {\it fixed configuration} of the projectile sources $\rho(x)$ and target fields $S(x)$, the function $\sigma(k)$ as a function of momentum has a maximum at some value ${\bf k}={\bf q}$.
Therefore the product in eq.(\ref{sigma}) is maximal for ${\bf k}={\bf p}={\bf q}$. The value of the vector ${\bf q}$ of course differs from one configuration to another, but the fact that momenta $\bf k$ and $\bf p$ are parallel does not .     
 Therefore after averaging over the ensemble $d^2N/dkdp$ has maximum at relative zero angle between the two momenta (as we have explained above, there is actually the second degenerate maximum at relative angle $\Delta\phi=\pi$).
The strength of the maximum of course depends on the detailed nature of the field configurations constituting the two ensembles (the projectile and the target).

Thus, angular correlations emerge as a result of target/projectile averaging procedure over  isotropic ensembles  of anisotropic  configurations.

We note that eq.(\ref{sigma}) holds in the case when one of the colliding objects is dense and another one is dilute\cite{baier},\cite{multigluons},\cite{jamal},\cite{pomloops2}. This is most likely not quite the situation encountered in the high multiplicity p-p events at LHC, where the density in the proton wave function is likely still not parametrically large, but is already not perturbatively small. The main features of our discussion are however born out by the expression eq.(\ref{product}) and we believe this approximation to be qualitatively correct. We note that the numerical calculations of \cite{ddgjlv} use a further perturbatively expanded version of eq.(\ref{sigma}).  The approach to the dense-dense regime has been developed in \cite{francois} and has been used in \cite{tuomas}. It is however not clear that in the LHC environment it is quantitatively more reliable than simple expression eq.(\ref{sigma}). When the dense system is produced in the final state, we expect the correlations produced by this mechanism to be washed out by the final state interactions and finally disappear for a very dense final state. There may be an alternative mechanism of producing angular correlations via radial flow effects\cite{nuclear ridge} which could be relevant to the ridge structure observed at RHIC\cite{ridgerhic}, but this is far beyond the scope of the present work.

Returning to eq.(\ref{product}), we note that angular correlations should be the leading $1/N_c$ effect \cite{LL,kl1,klw,KLc}. 
The leading $N_c$ piece in eq.(\ref{product}) comes from the configuration where the charge densities in each one of the single gluon production amplitudes are in the color singlet. The relevant average to calculate is
\begin{equation}
\langle \rho^a(x_1)\rho^a(\bar x_1)\rho^b(x_2)\rho^b(\bar x_2)\rangle_P
\langle{\rm Tr}\left\{[S^\dagger( x_1)-S^\dagger( z)]
                       [S(\bar x_1)-S(\bar z)]\right\}{\rm Tr}\left\{[S^\dagger( x_2)-S^\dagger( u)]
                       [S(\bar x_2)-S(\bar u)]\right\}\rangle_T\,.
                       \label{avera}
\end{equation}
On the target side one needs to calculate averages of observables of the type described in the large $N_c$ limit by the dipole model \cite{dipole} 
\begin{equation}
\langle{\rm Tr}\left\{[S^\dagger( x)S( z)]
                       \right\}{\rm Tr}\left\{[S^\dagger( y)S( u)]
                       \right\}\rangle_T=\langle s(x,z)s(z,x)s(y,u)s(u,y)\rangle_T\label{fad}
\end{equation} 
 where $s(x,y)=\frac{1}{N_c}{\rm Tr}[S^\dagger_F(x)S_F(y)]$ - is the  scattering amplitude of the fundamental dipole, and the equality in eq.(\ref{fad}) holds in the large $N_c$ limit.
 The approximation which is frequently used in the literature to calculate the averages of this type also invokes factorization
 \begin{equation}
 \langle s(x,y)\,s(u,v)\rangle=\langle s(x,y)\rangle\langle s(u,v)\rangle\,.
 \label{fact}
 \end{equation} 
 Strict factorization of the type eq.(\ref{fact}) is only possible if the statistical ensemble either consists of a single configuration, or the target fields on which the two dipoles scatter are completely independent of each other. There is of course no reason to expect that in the large $N_c$ limit fluctuations around some leading configurations are suppressed by powers of $1/N_c$. Likewise, since we are interested in dipoles which scatter within correlation radius of each other, the field configurations should be by definition correlated. Thus the factorization eq.(\ref{fact}) is not appropriate for study of correlations.
 
Our objective in this note is a pilot study of the evolution in energy of correlations which one can encode in the initial target ensemble.
As a framework for the evolution we take the projectile dipole model, which describes the evolution of $s$ in the leading $N_c$ limit. It is important at this point to avoid confusion and understand clearly which dipole model we are talking about, as there are two distinct approximations to the high energy evolution which are both sometimes called dipole model.

 In the first, which we will call target dipole model, one follows the evolution of the target wave function in terms of the density of dipoles (and their cumulants). This evolution as formulated in\cite{dipole} does not take into account finite density effects in the target wave function and is the large $N_c$ limit of the KLWMIJ evolution\cite{klwmij}. 
 Ref.\cite{yoshi} studied correlations and fluctuations of dipole density in this approach for a single dipole target and found them to be significant in the leading order in
  $1/N_c$. 
 
 Another dipole approach, which we will refer to as the projectile dipole model, evolves the projectile wave function according to dipole evolution. The projectile scattering amplitude is then calculated by approximating the scattering amplitude of each projectile dipole by an eikonal factor. This approximation can be reformulated as the evolution of the target wave function. In this form it is a large $N_c$ approximation to the JIMWLK evolution of the target wave function, that is, it does indeed take into account nonlinearities in the target evolution.

The two dipole approximations implement very different physics in the target wave function. Our choice of the projectile dipole approximation is motivated by the  expectation that high density effects in the target evolution should be important, and also by the fact that it is this approximation (or JIMWLK which includes $1/N_c$ corrections to it
 \cite{new}) that is used in current numerical studies of ridge\cite{ddgjlv}. We note however that this approximation {\it does not} take into account proper splittings of the target dipoles, as stressed in \cite{splitting}. As we show in the next section numerically, and explain qualitatively in Sec. 4 this deficiency turns out to be crucial in the inability of this approximation to correctly evolve correlations present in the initial ensemble.

The probability distribution of the dipole model $W[s]$ evolves with rapidity according to \cite{LL,kl1}
\begin{equation}
 {d\over d Y}W[s]=
 \frac{ \bar{\alpha}_s}{2\,\pi}\,
\int_{x,y,z}\frac{(x-y)^2}{(x-z)^2\,(z\,-\,y)^2}\,\left[\,s(x,\,y)\,-\,
\,s(x, z)\,s(y,z)\,\,\right]
\frac{\delta}{\delta s(x, y)} W[s]
\end{equation}
with $\bar \alpha$ - the 'tHooft coupling, which is finite at infinite $N_c$.
Our strategy is to choose an ensemble $W_0[s]$ of initial configurations $s(x,y)$,  which contains nontrivial angular correlations.  Each configuration of the ensemble is evolved independently according to the BK equation \cite{kl1}. The correlations at the final rapidity are then calculated by averaging the correlator over the ensemble of solutions $s_Y(x,y)$.
\begin{equation}
\int Ds W_Y[s]s(x,y)s(u,v)= \int Ds W_0[s]s_Y(x,y)s_Y(u,v)
\end{equation}
where $s_Y(x,y)$ is the solution of the BK equation with initial condition $s(x,y)$.
 
This procedure is similar to the one implemented in \cite{nestor}.  However, the focus of \cite{nestor} was in fluctuations of the saturation scale and thus all configurations in the initial ensemble in \cite{nestor} were chosen to be isotropic. In order to study  angular correlations we have to allow the individual members $s(x-y)$ of the initial ensemble to be anisotropic. The rotational invariance is restored by averaging over the whole ensemble rather than configuration by configuration.
 
Here we report on our initial results, which mostly aim at qualitative  understanding of rapidity dependence of angular correlations within the projectile dipole evolution.  Like in most BK studies we do not consider impact parameter dependence. We view the resulting correlations as correlations at a fixed impact parameter, thus strictly speaking study of correlations in impact parameter plane are beyond our current calculation. Nevertheless we do not expect our results on weakening of correlations with rapidity at a fixed impact parameter to be affected by configuration by configuration fluctuations in the impact parameter plane.

For the sake of simplicity we do not calculate the two gluon production rate eq.(\ref{product}) but rather examine the simplest observable that can exhibit angular correlations - the correlator of two dipole scattering amplitudes $s(x,y)s(u,v)$.

 \section{BK equation, initial conditions, and angular dependence} 

\subsection{BK equation and initial conditions}
The BK equation for the imaginary part of the dipole scattering amplitude $N(\vec r) =1-s(\vec r)$ (assuming impact parameter independent configurations) is:
\beq\label{BK}
\partial_Y\,N(\vec r)\,=\, \frac{ C_F\,{\alpha}_s}{2\,\pi}\,\int d^2 \vec r^\prime {\vec r^2\over  \vec r^{\prime\,2}\, (\vec r-\vec r)^2}\ [N(\vec r^\prime)\,+\,N(\vec r-\vec r^\prime) \,-\, N(\vec r)\,-\,N(r^\prime)\,N(\vec r-\vec r^\prime)]\,.
\eeq
Here $\vec r=\vec x-\vec y$ is a vector (in the transverse plane) connecting the two legs of the dipole, and  $r=|\vec r|$. 

As discussed above, we have to specify the initial ensemble of configurations of $N(\vec r)$. We choose all the configurations of the ensemble to have similar radial dependence, but distribute them homogeneously with respect to angle $\theta$. A representative configuration at some initial rapidity $Y_0=\ln 1/x_0$ is taken as
\beq\label{ini}
N(Y_0,\vec r)\,=\,1\,-\,Exp\Bigg\{\,-\,a\,r^2\,xg^{LOCTEQ6}(x_0,4/r^2)\,F(\theta)\Bigg\};\ \ \ \ \ \ \ \ \ \ \ \ a\,=\,{\alpha_s(r^2)\,\pi \over 2\,N_c \,R^2}\,.
\eeq
Apart from the angle dependent function $F$, this is the same initial condition as used in \cite{GLLM} to fit HERA data. In \cite{GLLM}, the $F_2$ low $x$ data were reproduced using the BK
equation with running coupling \footnote{In \cite{GLLM} the running coupling was taken to depend on the external dipole size.  We stick to this choice also in the present work,
even though there exist a more rigorous  way to introduce the running into the BK equation\cite{KW}. These fine differences are immaterial for the qualitative features of our results.}. 
The parameters used in \cite{GLLM} are: $x_0=10^{-2}$ and the effective proton's radius $R$ fitted to the $F_2$ data, $R^2=3.1\,(GeV^{-2})$. 

\begin{figure}
\begin{tabular} {cc}
\includegraphics[width=7cm]{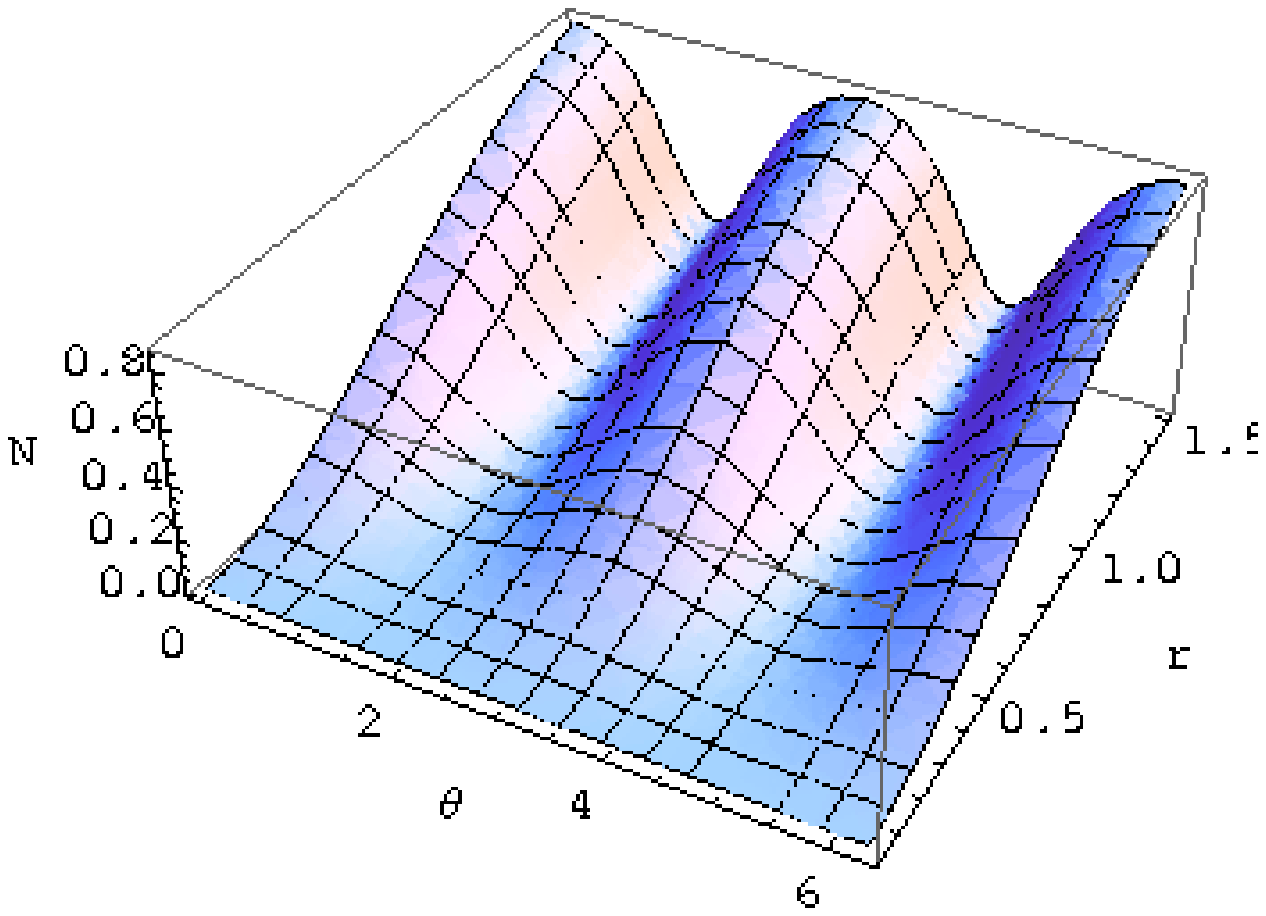} \ \ \ \ \ \ \ \ \ & \ \ \ \ \ \ \ \
\includegraphics[width=5cm]{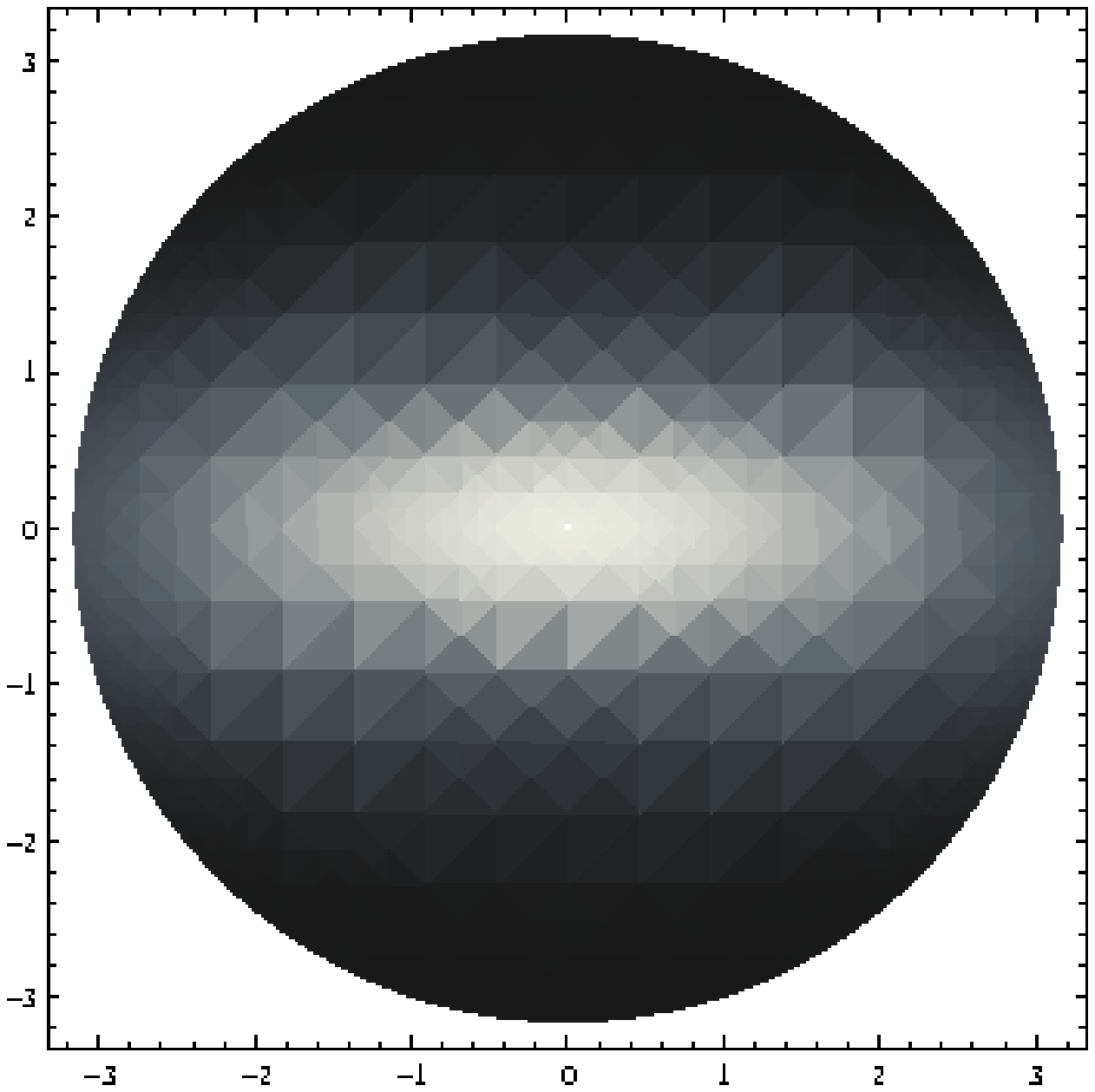} \ \
\end{tabular}
\caption{Initial conditions (\ref{ini}). Left: profile of $N$ as a function of $r$ and $\theta$. Right: the same in polar coordinates.}
\label{inifig}
%\end{figure}
\end{figure}

The function $F(\theta)$ takes into account angular modulations of the scattering amplitude relative to some axis thereby reflecting anisotropy
of a given target field configuration.  For our study we choose  $F(\theta)$ of the form:
\beq\label{F}
F(\theta)\,=\, {1\over 4}\,+\,{3\over 2}\,cos^2( \theta)\,.
\eeq
In contrast to previous numerous studies of the BK equation, in which the initial conditions were parameterized with respect to the dipole size only,   eq. (\ref{ini}) provides 2-dimensional
initial data set (Fig. \ref{inifig}; For all the plots $r$ is given in GeV$^{-1}$).  

The $cos^2(\theta)$ dependence is motivated by dipole interaction with a constant target chromo-electric field $\vec E$, eq.(\ref{amplitude}). In fact, even though this is quantitatively not quite true,  qualitatively the scattering amplitude (\ref{ini}) can be thought of as the scattering amplitude on a fixed, constant chromo-electric field configuration in the target.
The function $F(\theta)$ has period $\pi$ and is symmetric under $\theta\rightarrow \pi-\theta$. For this reason in what follows we will be quoting only results in the first quarter. 

%\begin{figure}

%In the right plot of the figure \ref{inifig}, one can think of one leg of the dipole as being placed in the origin while another one probing 
%the target field at a distance $r$ from the center. One, however, should not be  confused as this picture does not represent a profile in the impact parameter space.
%All secondary dipoles are produced at the same impact parameter.

We now define the ensemble of initial conditions by homogeneously distributing the direction of the target field in the impact parameter plane.
%As we have argued above, we have to average over an ensemble of initial conditions. Moreover, the ensemble has to restore the overall rotational symmetry.  Any single
%initial configuration of the form (\ref{ini}) has a preferable axis. It is thus natural to have an ensemble in which the  direction of this axis is a random variable.
%It is easy to introduce such an ensemble.  
In practical terms this amounts to shifting the angle $\theta$ in eq.(\ref{F}) by another angle $\delta$, which is taken to be a random variable with the probability 
distribution $W[\delta]=1/2\pi$, constant  for any $\delta$ ranging from $0$ to $2\pi$. Averaging over such ensemble restores rotational invariance.  In particular, for example
\beq\label{Fav}
\langle F \rangle_\delta\,=\,\int_0^{2\pi} d\delta \,F(\theta+\delta)\,W[\delta]\,=\,1\,.
\eeq
Below, we will consider various observables, such as $\langle N(Y,r,\theta,\delta)\rangle_\delta $ as well as two-dipole correlator
$\langle N(Y,r_1,\theta_1,\delta)\,N (Y,r_2,\theta_2,\delta)\rangle_\delta$ averaged with respect to $\delta$ with the weight  $W[\delta]$. 

\subsection{Single configuration solution} 

\begin{figure}
\begin{tabular} {ccc}
\includegraphics[width=5.7cm]{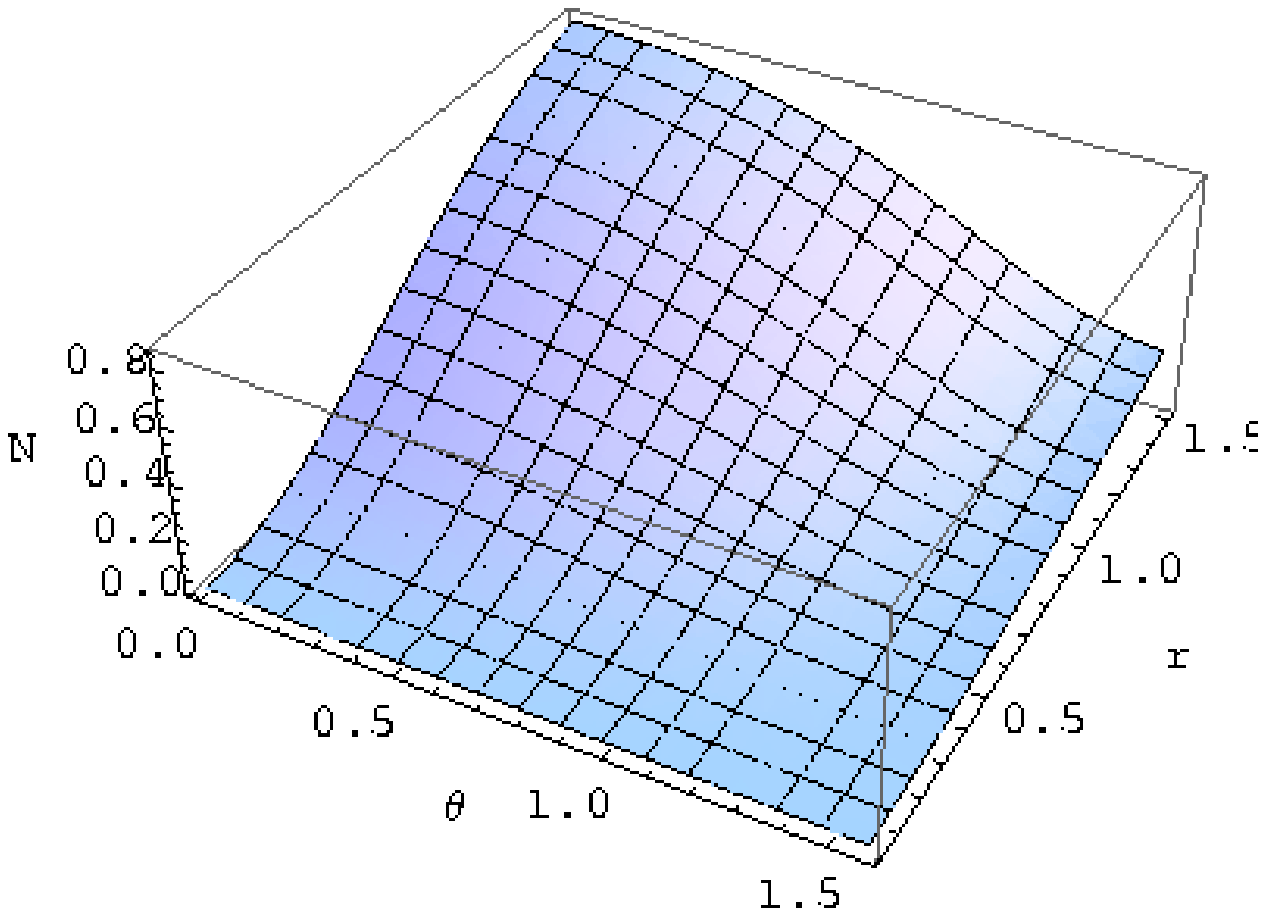} \    &  \ 
\includegraphics[width=5.7cm]{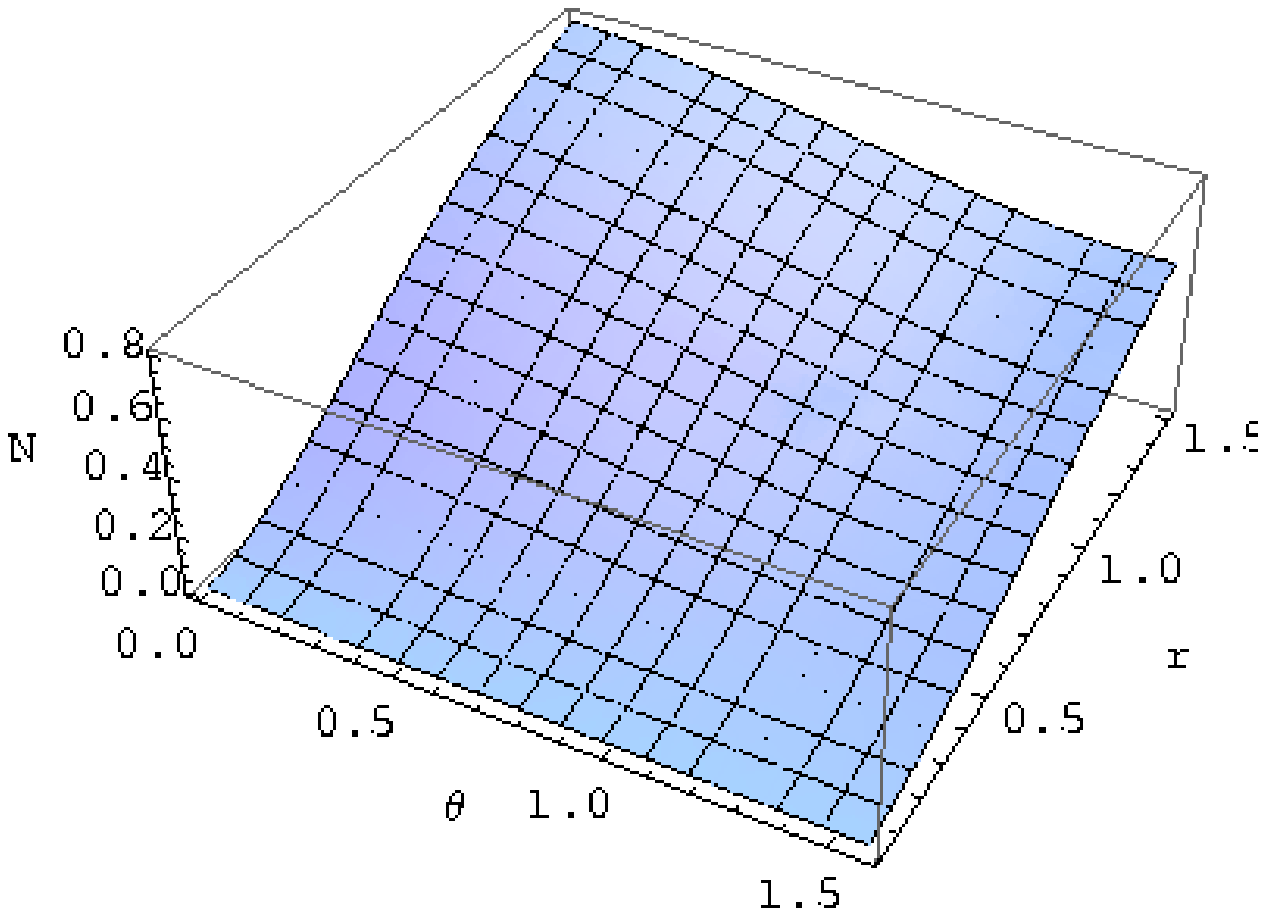}\    &    \  
\includegraphics[width=5.7cm]{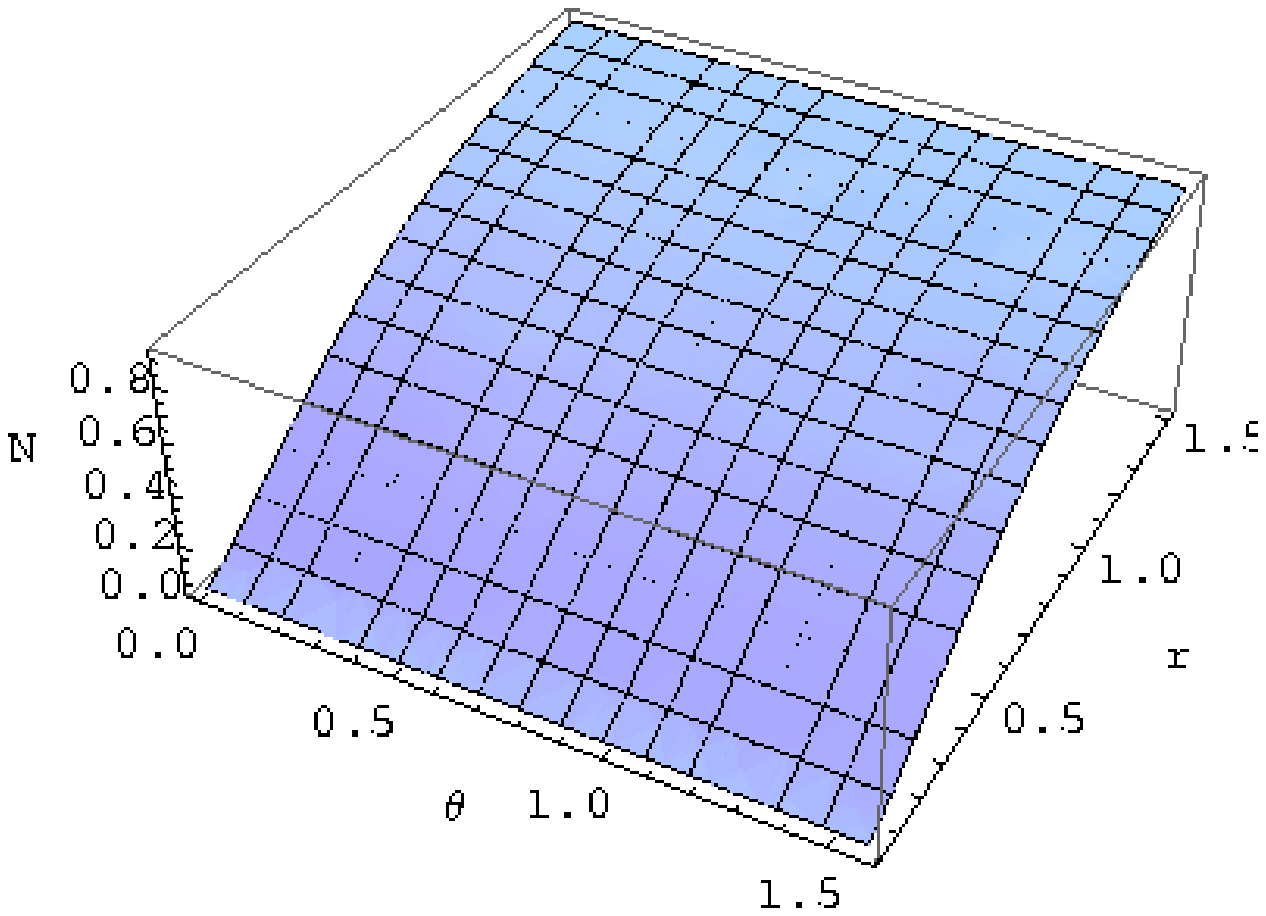} \ \
\end{tabular}
\caption{$N$ as a function of $r$ and $\theta$ at various values of rapidity: $Y=Y_0\simeq 4.6$, $Y=6$, $Y=10$}
\label{inifig1}
\end{figure}

In Fig.  \ref{inifig1} we present our numerical solution of the BK equation with the initial condition (\ref{ini},\ref{F}). 

The main qualitative feature of the evolution is quick isotropisation even on a single initial configuration, without the ensemble averaging.
One way to quantify the effect is to focus on the saturation scale, defined in a standard manner
$$N(Y,R_S,\theta)\,=\,1/2\,.$$
The resulting saturation radius $R_s$ is now both rapidity and angle dependent.  
\begin{figure}
\centerline{\includegraphics[width=7.5cm]{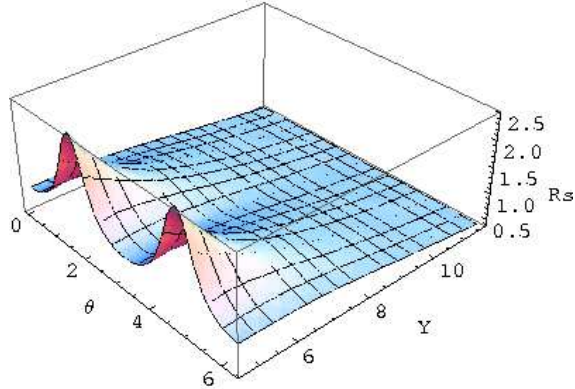}} 
\caption{Saturation radius as a function of angle and rapidity.}
\label{rs}
\end{figure}
The initial strong angular dependence of $R_S$ completely disappears after evolution by about five units of rapidity (Fig. \ref{rs}).

\begin{figure}
\begin{tabular} {cc}
\includegraphics[width=7cm]{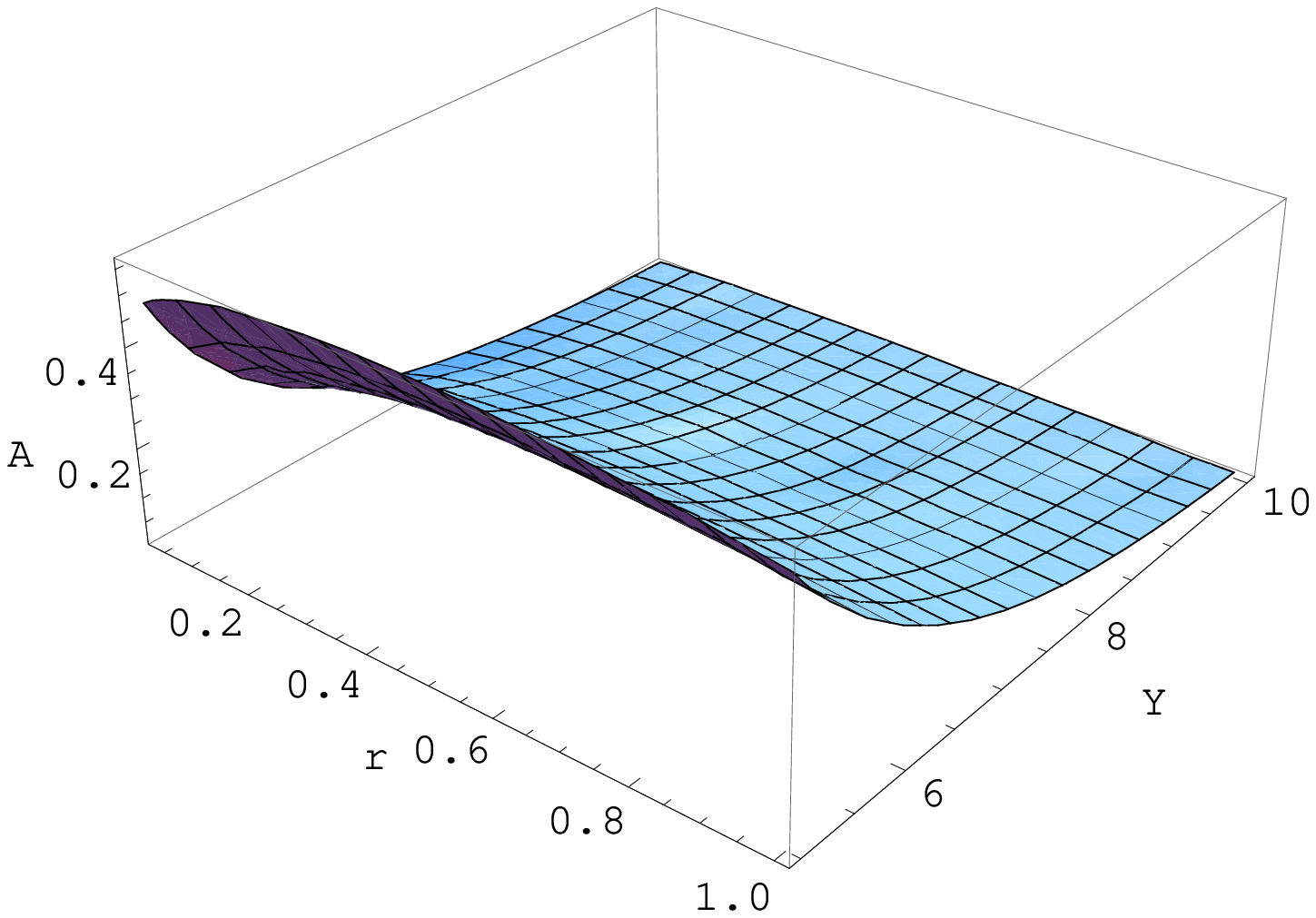} \ \ \ \ \ \ \ \ \ & \ \ \ \ \ \ \ \
\includegraphics[width=7cm]{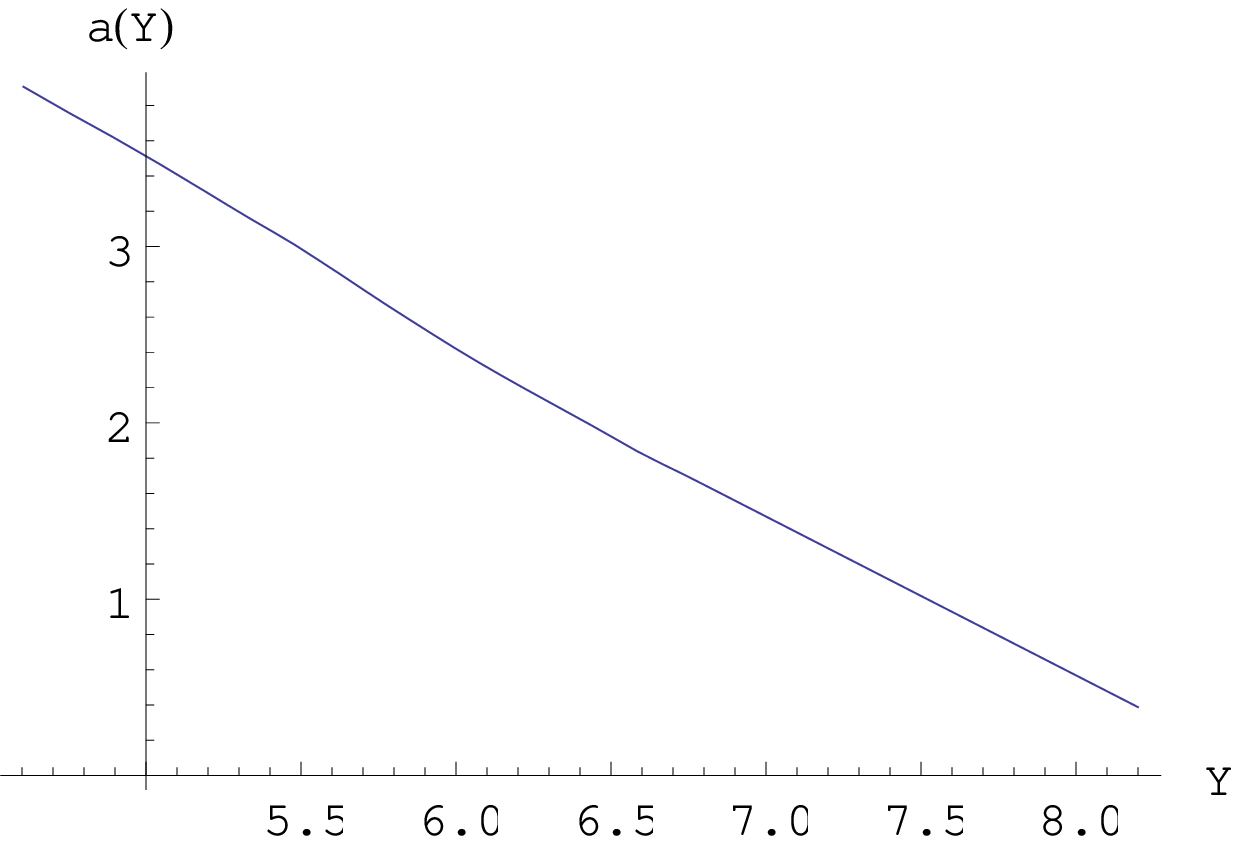} \ \
\end{tabular}
\caption{a(Y) and A(Y)}
\label{inifig2}
\end{figure}

In Fig.(\ref{inifig2}) we plot another measure of anisotropy
\beq\label{A}
A(Y,r)\,\equiv\,{N(Y,r,0)\,-\,N(Y,r,\pi/2)\over N(Y,r,0)\,+\,N(Y,r,\pi/2)}
\eeq
 as a function of $Y$ and $r$. Again we observe an exponentially fast disappearance of the anisotropy.  
 
 Fig.(\ref{inifig2}) has a curious feature, which may or may not be important. The anisotropy $A$ seems to be maximal at a fixed scale $r\simeq r_{max}=0.5\,GeV^{-1}$ independently of rapidity. The origin of this scale is not clear to us and it may just be a numerical accident related to the form of our initial configuration. We have checked however that $A$ remains maximal at $r\simeq r_{max}$ even when $R$ of the initial condition is varied, thus possibly hinting at another origin. At any rate one can follow the weakening of anisotropy by following the ratio $A$ at $r_{max}$. We find
\beq A(r_{max})\,\sim\, e^{-\lambda_A\,Y}\,, \ \ \ \ \  \ \ \ \ \ \ \ \ \ \ \ \ \lambda_A\,\simeq\,0.6\,.
\eeq 
We also fit in the pre-saturation regime $N(Y,r_{max},0)\sim e^{0.2\, Y}$ while $N(Y,r_{max},\pi/2)\sim e^{0.4\, Y}$\footnote{Note that the exponential behavior in these "`fits"' is only approximate, and is simply given to guide the eye.}.

We have also looked at the rate with which the scale of fixed anisotropy shrinks with rapidity. We follow the scale $r$ at which $A$ takes a constant value, say 10\%.
Solving for $A(Y,r)=0.1$ leads to the scale $r=a(Y)$. The scale $r$ moves towards smaller dipole sizes with what looks like a constant rate.
Our fit gives  $a(Y)\simeq 8.5-Y$. Note, however, that the fit is limited to not very high rapidities. At larger rapidities 
the angular dependence of $N(\vec r)$ is washed away and the anisotropy drops below 10\% level for all dipole sizes. This is clearly seen on the $A$ plot.

It is important to stress that the wash-away of angular anisotropy occurs even for very small dipole sizes, where the evolution is governed entirely by
the BFKL dynamics. The mechanism behind  this fast isotropisation therefore must be rapid angular decorrelation of emitted gluons inside the BFKL ladder. 

\subsection{Averaged fluctuations}

\begin{figure}
\centerline{\includegraphics[width=7cm]{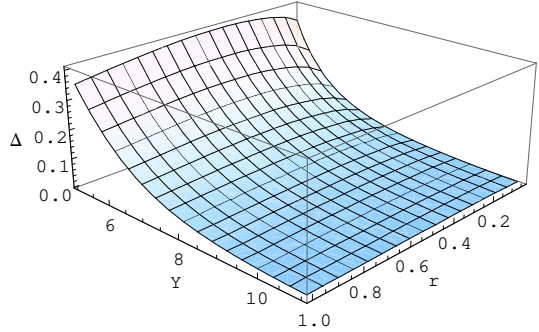}} 
\caption{The angle independent correlator $\Delta$.}
\label{fig6}
\end{figure}

Moving on to observables averaged over the whole ensemble of initial conditions, we first plot the fluctuation of the simplest angle independent correlator (Fig. \ref{fig6}).  
\beq
\Delta(Y,r)\,\equiv\,{\sqrt{\langle N(Y,r,\theta,\delta)^2\rangle_\delta\,-\,\langle N(Y,r,\theta,\delta)\rangle_\delta^2}\over \langle N(Y,r,\theta,\delta)\rangle_\delta}\,.
\eeq 

\begin{figure}
\begin{tabular} {cc}
\includegraphics[width=6cm]{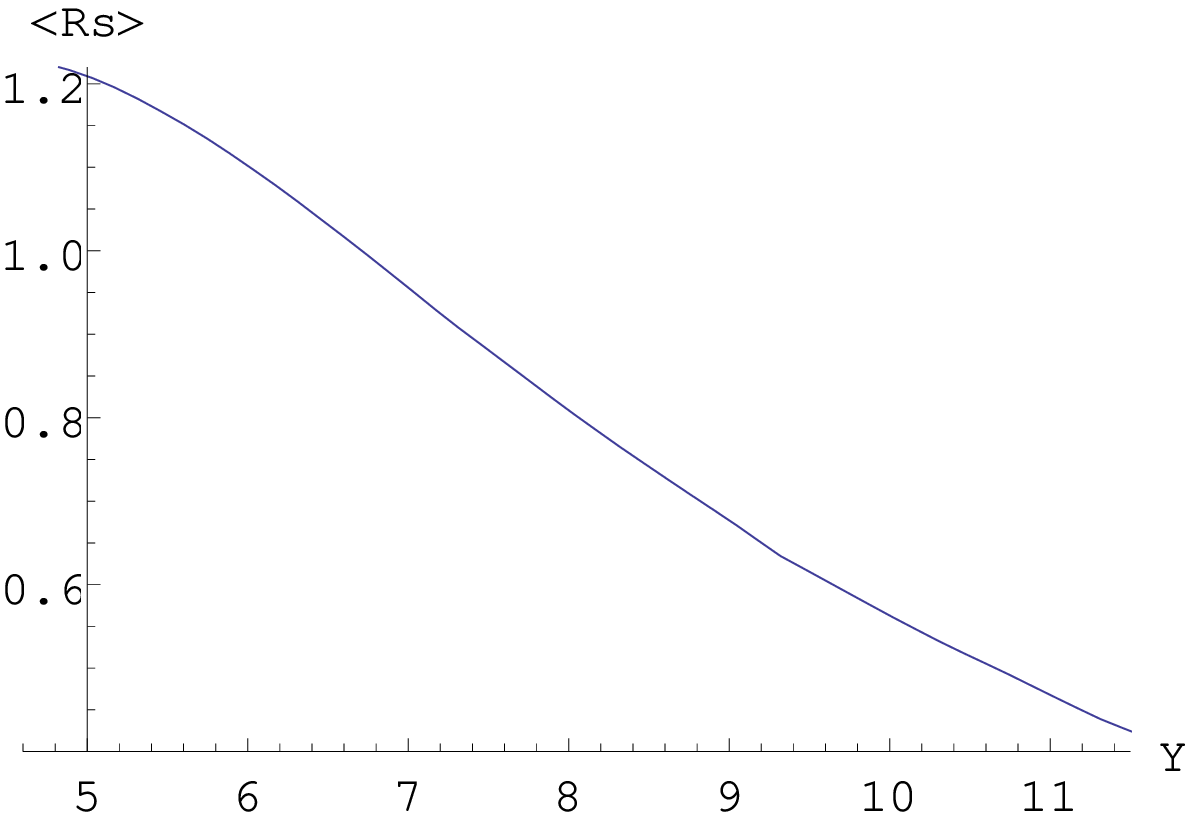} \ \ \ \ \ \ \ \ \ & \ \ \ \ \ \ \ \
\includegraphics[width=6cm]{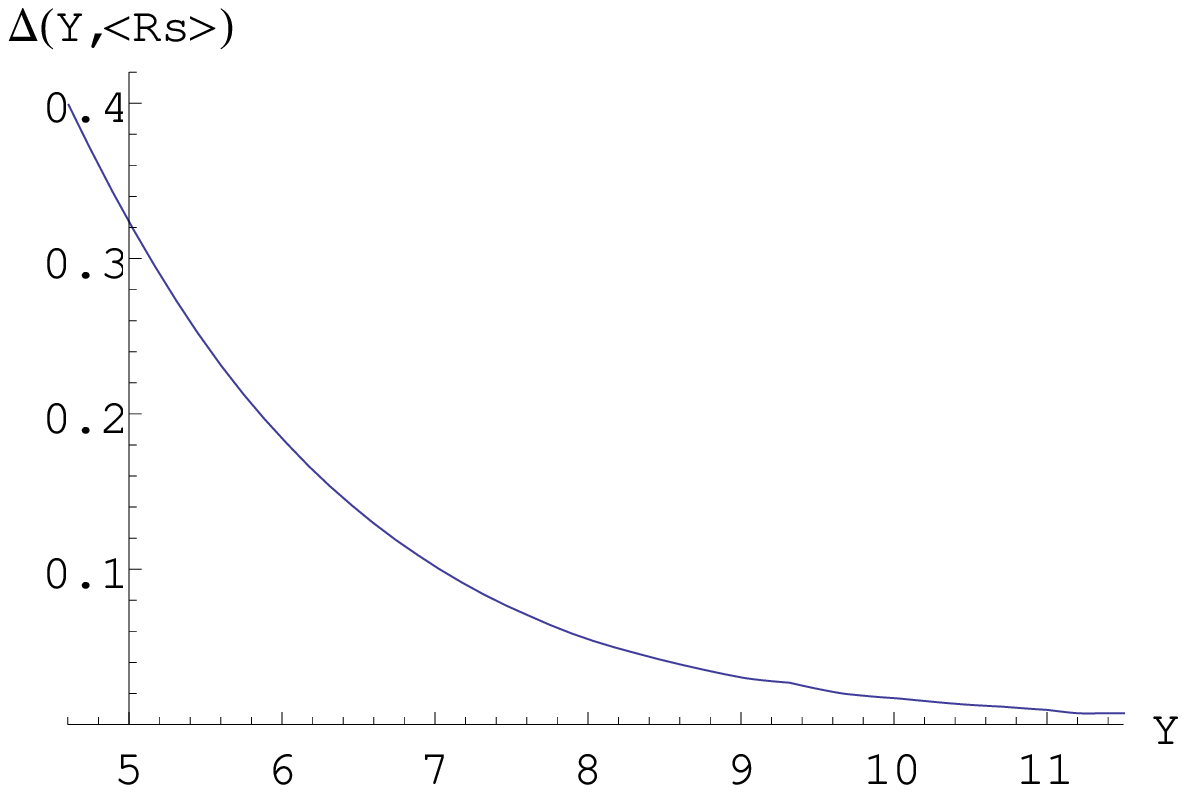} \ \
\end{tabular}
\caption{Averaged saturation radius and the fluctuation $\Delta(Y,\langle R_s\rangle$)}
\label{inifig3}
\end{figure}

We again observe the appearance of a rapidity-independent maximum at the scale $r_{max}$, although this maximum is quite shallow. The fluctuation $\Delta$ decays exponentially fast with rapidity with the same
exponential $\lambda_A$:
\beq
\Delta(Y,r_{max})\,\sim\, e^{-\lambda_A\,Y}\,.
\eeq
We have found that the exponent $\lambda_A$ emerges in several other observables we looked at.
Defining the angular-averaged saturation scale $\langle R_s\rangle$  and plotting the fluctuation $\Delta$ at the scale $\langle R_s\rangle$ (Fig. \ref{inifig3})
we again find
\beq
\Delta(Y,\langle R_s\rangle)\,\sim\, e^{-\lambda_A\,Y}\,.
\eeq
and conclude that all fluctuations are rapidly washed away at the saturation scale.

\subsection{Angular Correlations}

\begin{figure}
\begin{tabular} {cc}
\hspace*{-2cm}\includegraphics[width=11cm]{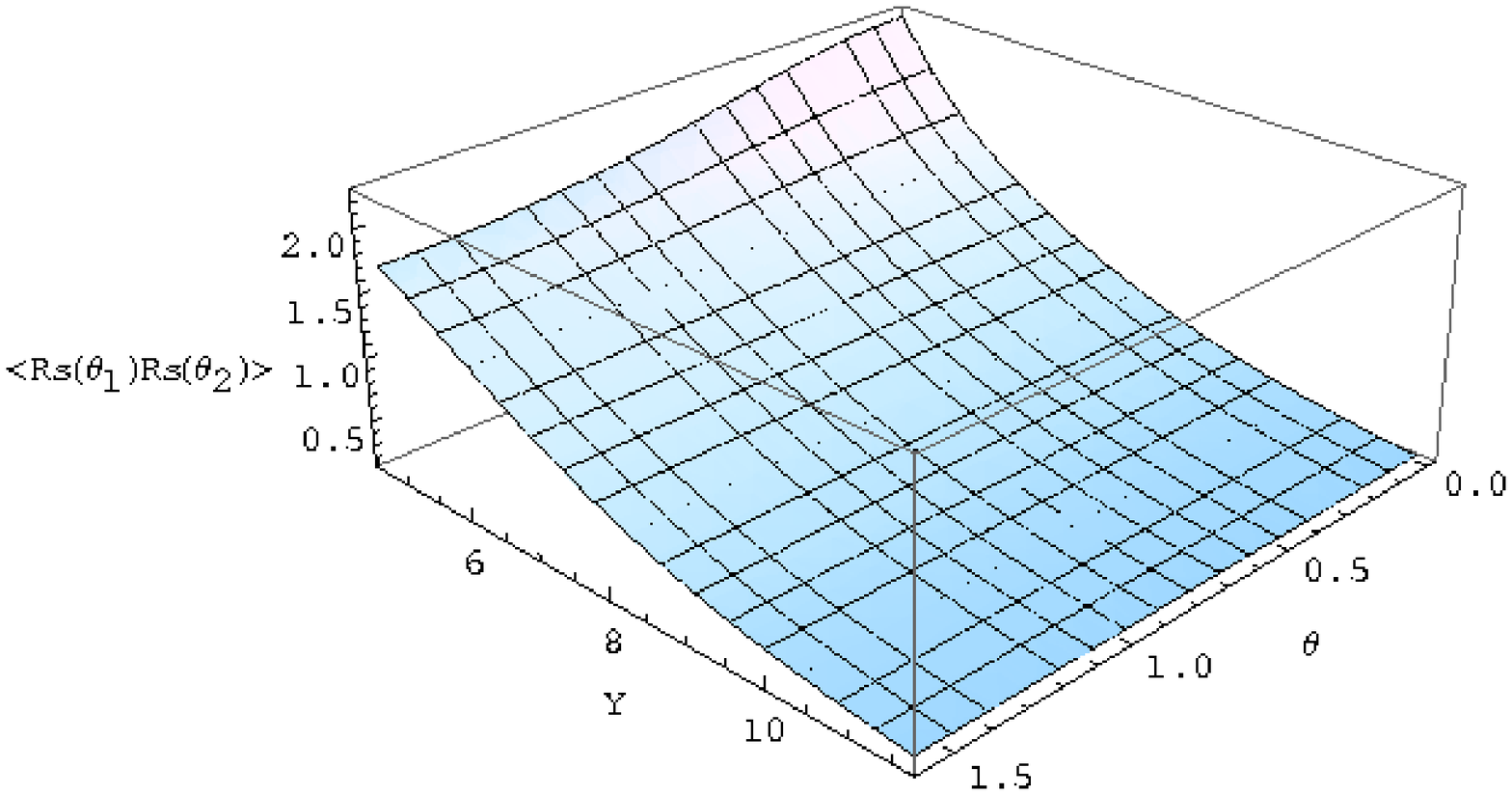}  & 
\includegraphics[width=7cm]{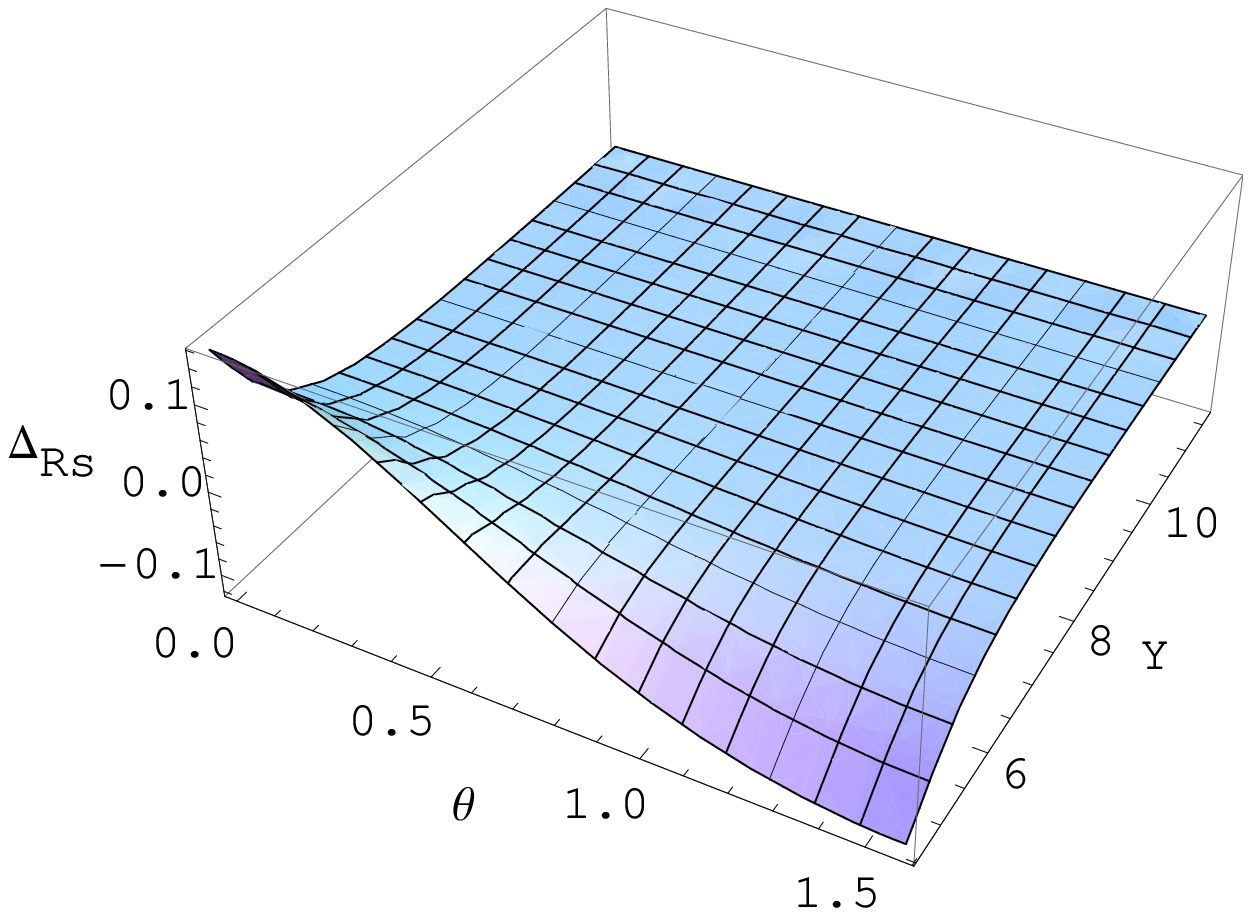} \ \
\end{tabular}
\caption{Left:Angular correlations of the saturation radius. Right: Normalized correlations}
\label{inifig4}
\end{figure}

Finally, we looked at angle dependent fluctuations of the dipole amplitude and related quantities.
The first quantity we plot (Fig. \ref{inifig4}) is the correlator of two saturation scales $\langle R_s(\theta_1)R_s(\theta_2)\rangle_\delta$ and 
the normalized correlation of the saturation radii
\beq
\Delta_{R_s}(Y,r,\theta)\,\equiv\,{\langle R_s(Y,\theta_1,\delta)\,R_s(Y,\theta_2,\delta)\rangle_\delta\,-\,\langle R_s(Y,\theta_1,\delta)\rangle_\delta\,\langle R_s(Y,\theta_2,\delta)\rangle_\delta\over \langle R_s(Y,\theta_1,\delta)\rangle_\delta^2}\,,\ \ \ \ \ \  \ \ \ \ \ \ \ \ \ \ \ \ \theta=\theta_1-\theta_2\,.
\eeq

Both quantities become angle independent when evolved by about five units of rapidity.

\begin{figure}
\begin{tabular} {cc}
\includegraphics[width=8cm]{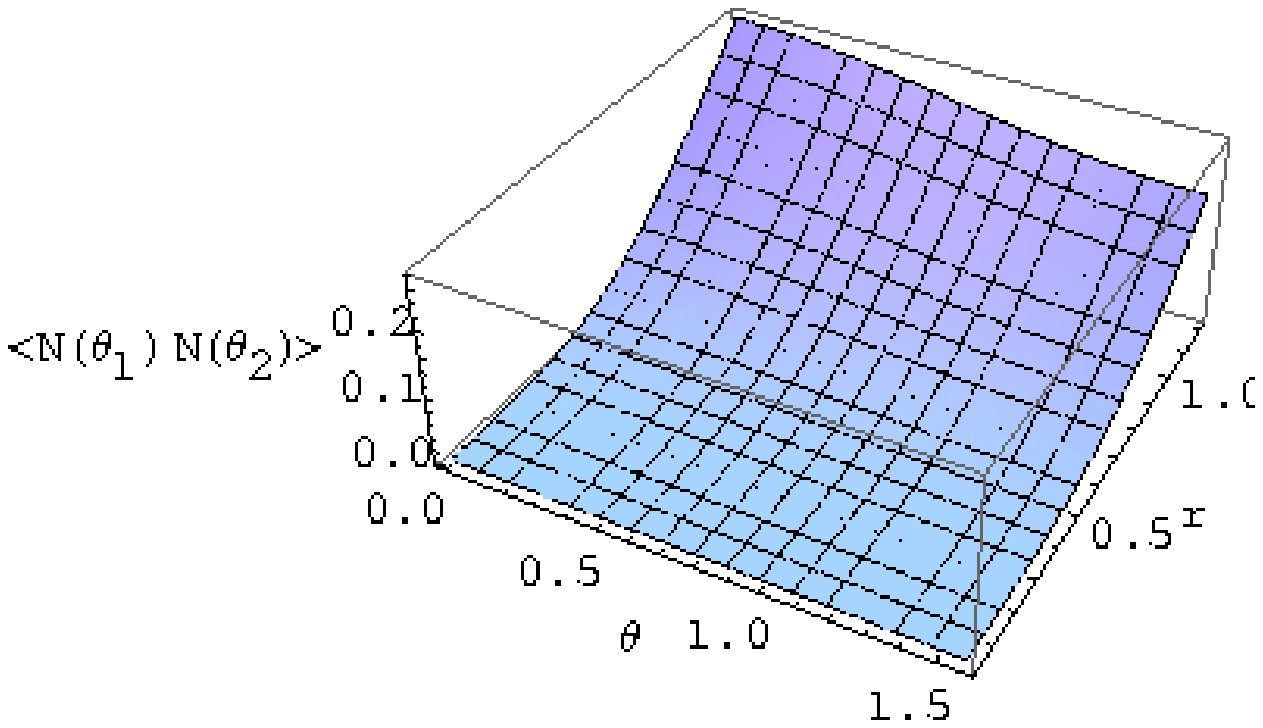}     &  
\includegraphics[width=8cm]{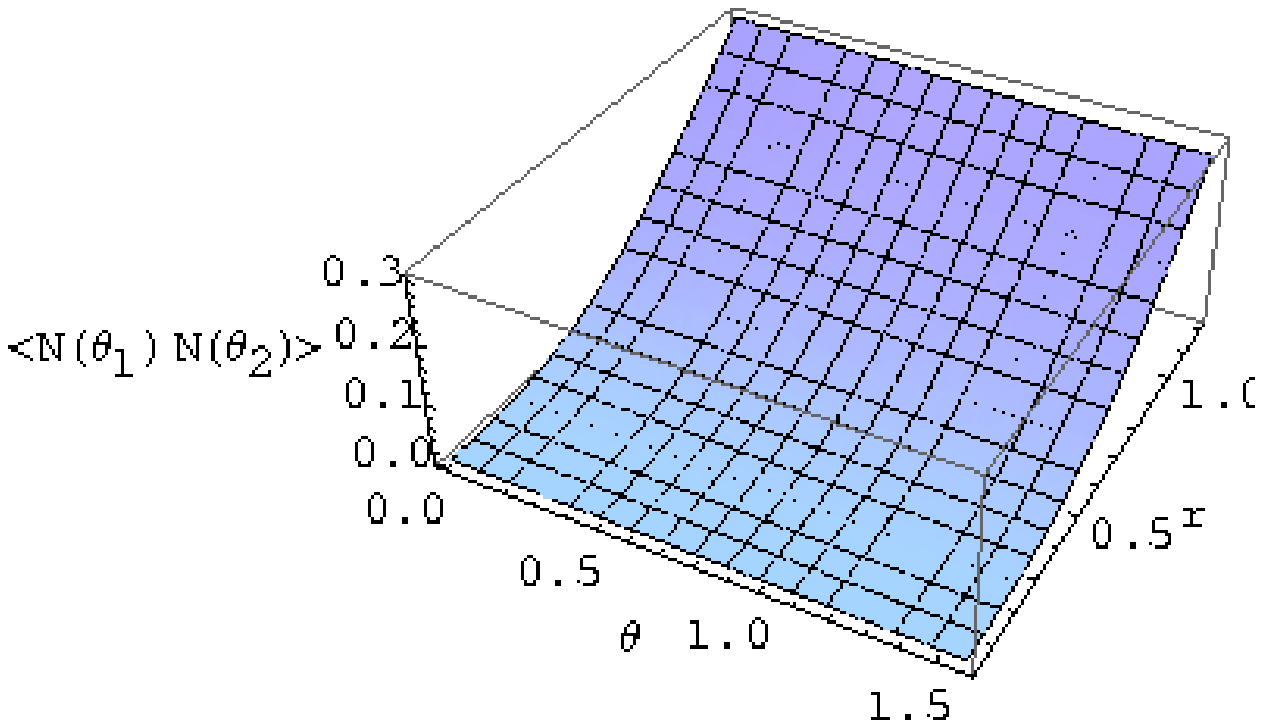} \ \
\end{tabular}
\caption{Angular correlations of $N$. Left: $Y=Y_0\simeq 4.6$; Right: $Y=6$}
\label{inifig5}
\end{figure}

\begin{figure}
\begin{tabular} {ccc}
\includegraphics[width=6cm]{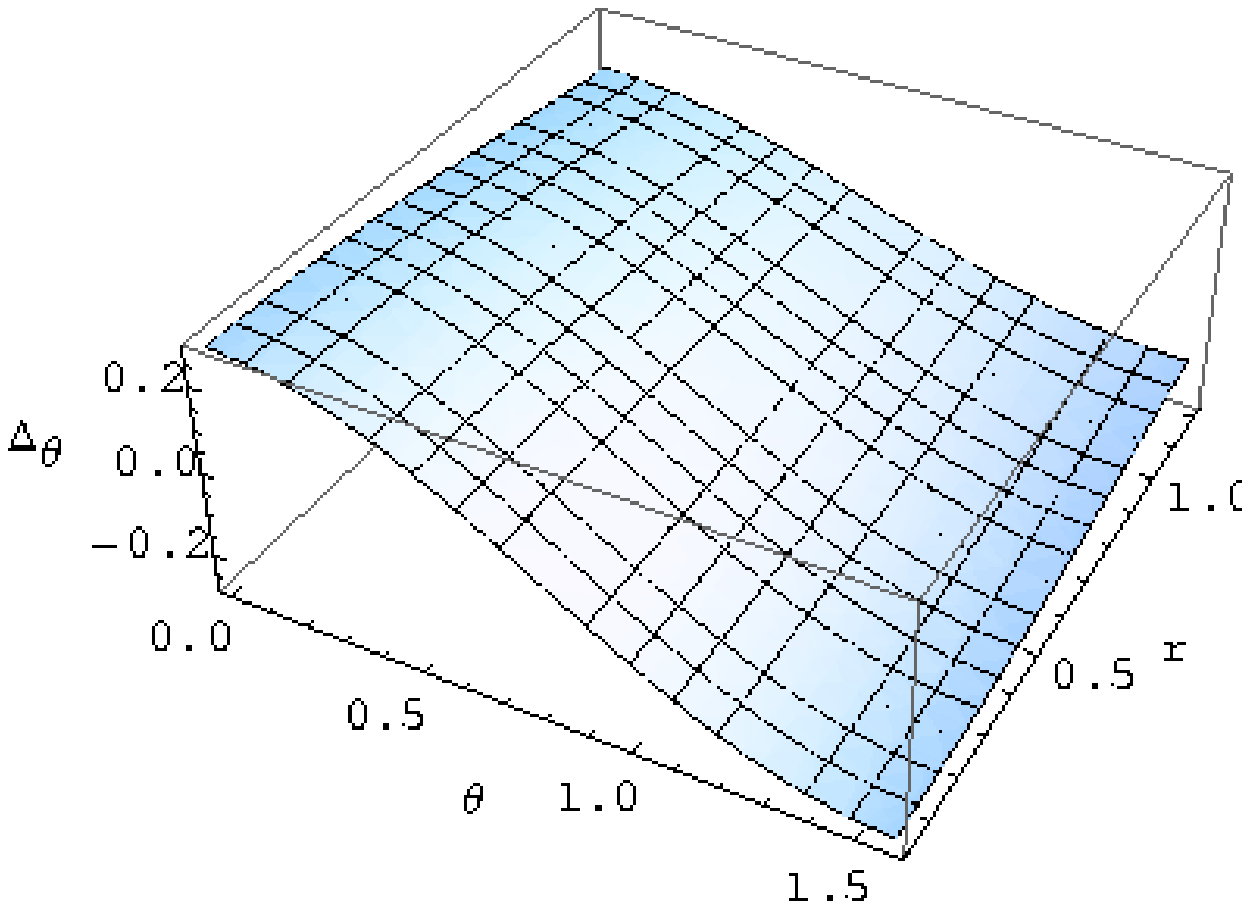}     &  
\includegraphics[width=6cm]{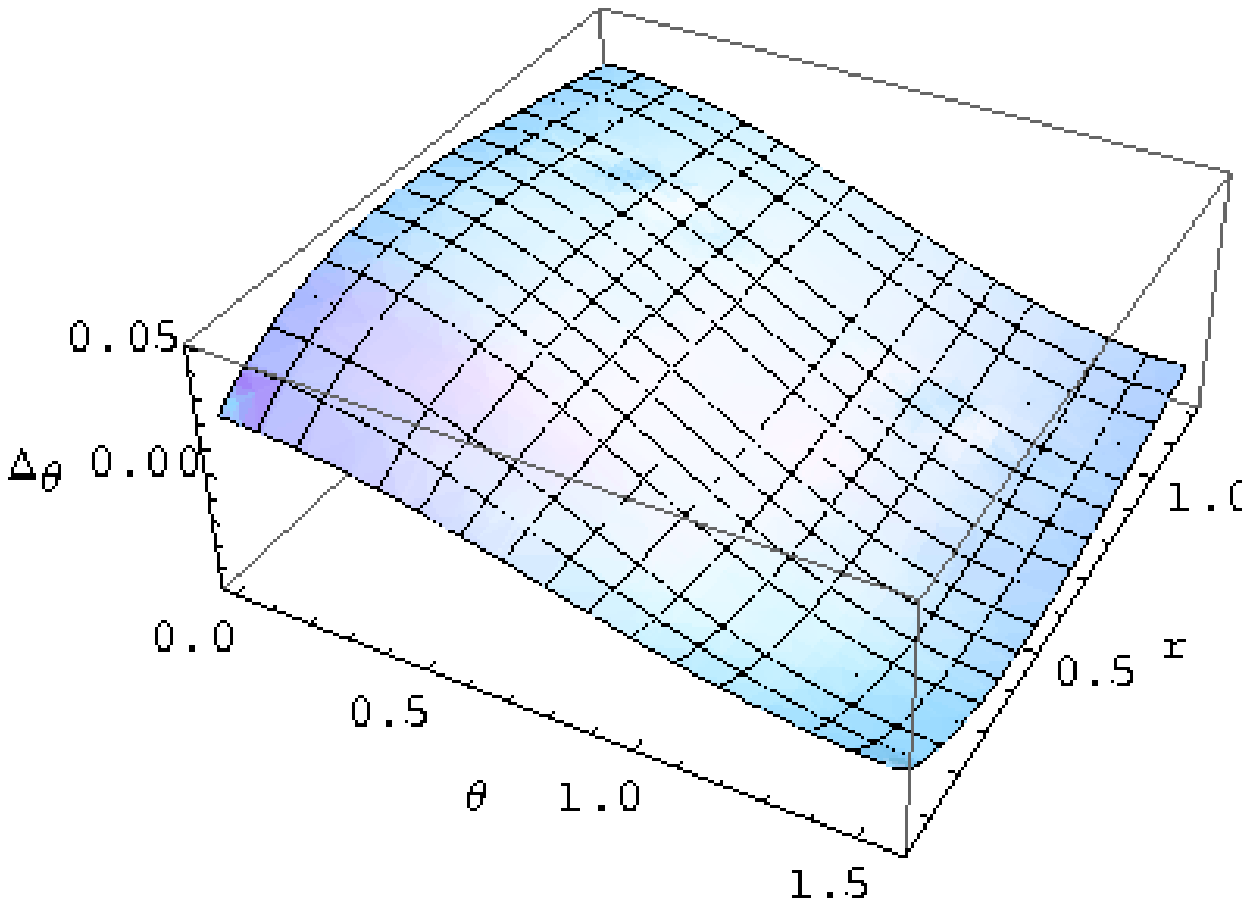}  &
\includegraphics[width=6cm]{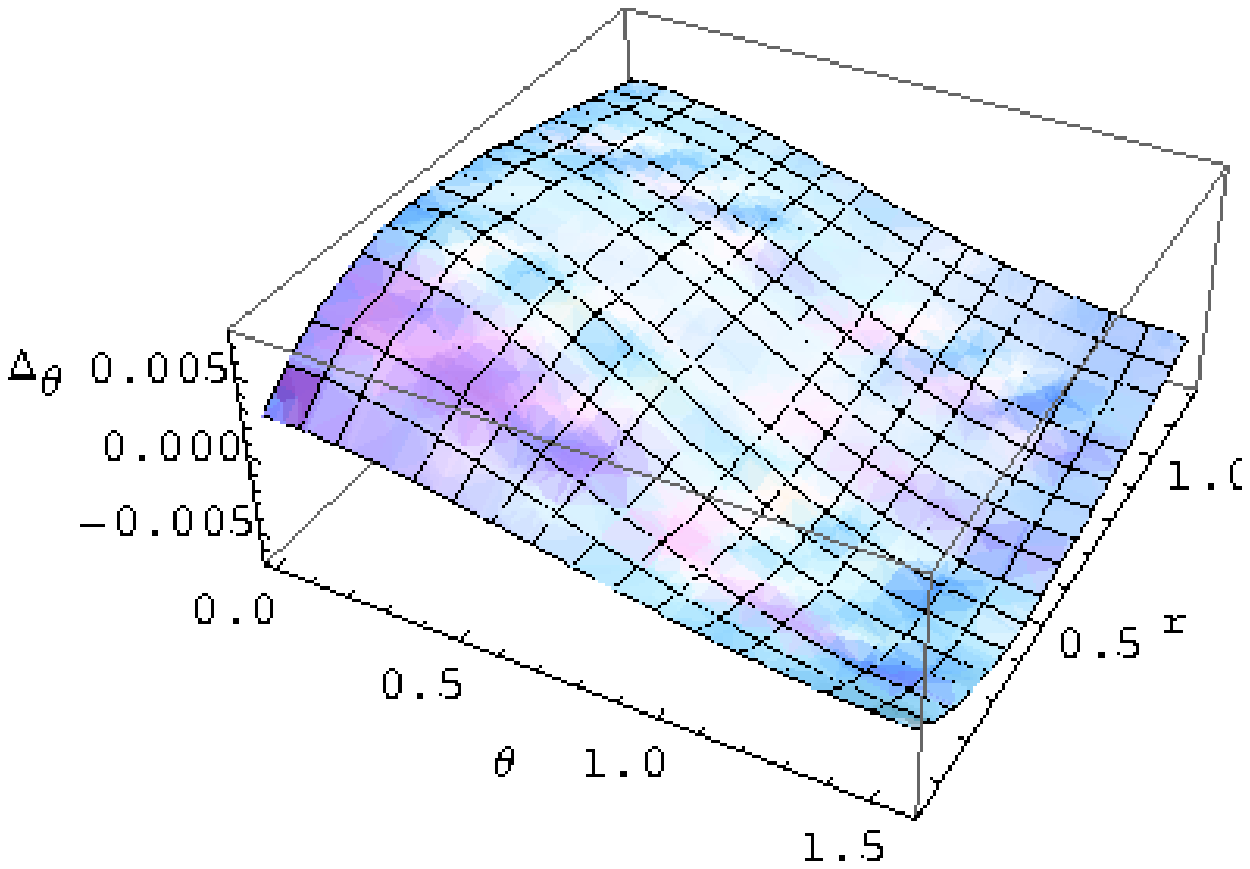}
\ \
\end{tabular}
\caption{Normalized angular correlations $\Delta_\theta$. Left: $Y=Y_0\simeq 4.6$; Middle: $Y=6$; Right: $Y=7.5$}
\label{inifig6}
\end{figure}

The angular correlations of the dipole amplitude itself behave in a similar fashion. We plot  (Fig. \ref{inifig5})
the correlator $\langle N(Y,r,\theta_1)\,N(Y,r ,\theta_2)\rangle_\delta$.
and the normalized fluctuation (Fig. \ref{inifig6})\footnote{We plot angular correlations only for dipoles of the same size. The picture is qualitatively the same also for different size dipoles.}
\beq
\Delta_\theta(Y,r,\theta)\,\equiv\,{{\langle N(Y,r,\theta_1,\delta)\,N(Y,r,\theta_2,\delta)\rangle_\delta\,-\,\langle N(Y,r,\theta_1,\delta)\rangle_\delta\,\langle N(Y,r,\theta_2,\delta)\rangle_\delta}\over \langle N(Y,r,\theta_1,\delta)\rangle_\delta^2}\,,\ \ \ \ \ \  \ \ \ \ \ \ \ \ \ \ \ \ \theta=\theta_1-\theta_2
\eeq

%We once again observe the emergence of the scale $r_{max}$
The normalized fluctuation decreases with rapidity approximately as
$$\Delta_\theta(Y,R_s(Y),\theta)\sim e^{-2\,\lambda_A\,Y}\,.$$

\begin{figure}
\begin{tabular} {cc}
\includegraphics[width=8cm]{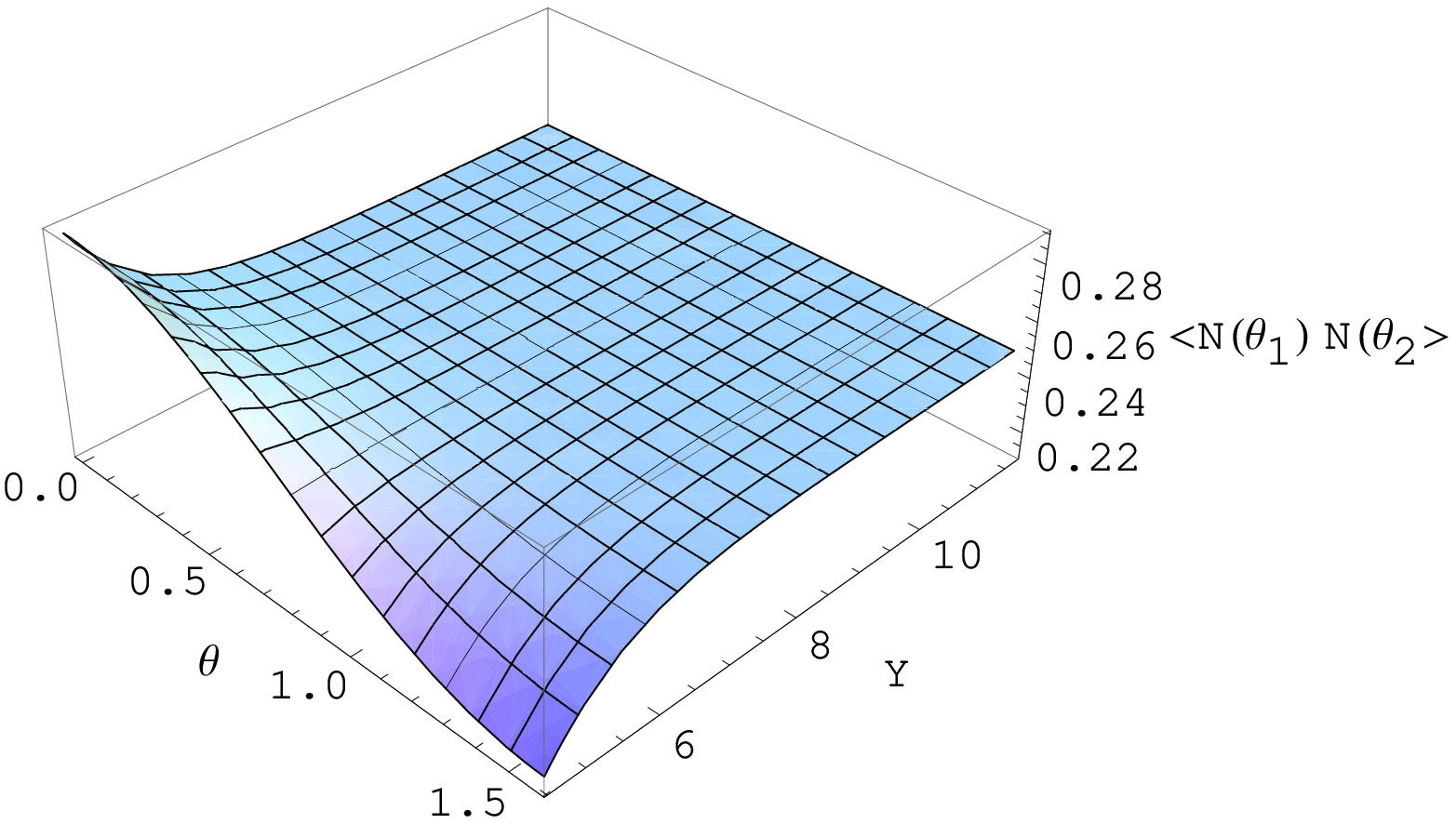} \ \    &  \ \
\includegraphics[width=7cm]{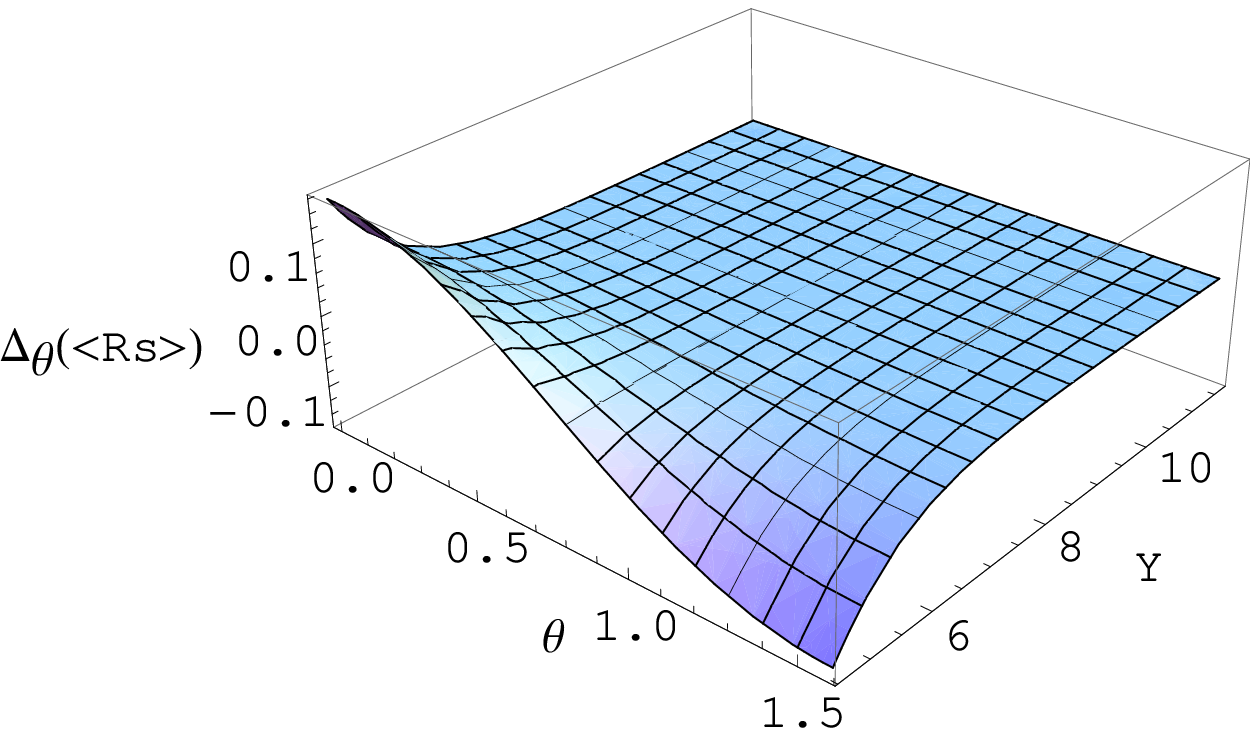} \ \
\end{tabular}
\caption{Angular correlations at the saturation scale $R_s(Y)$. Left: $\langle N(Y,R_s(Y),\theta_1) N(Y,R_s(Y),\theta_2\rangle) $; Right: $\Delta_\theta(Y,R_s(Y),\theta)$}
\label{inifig7}
\end{figure}

Finally we plot the normalized correlation at the saturation scale as a function of rapidity (Fig. \ref{inifig7}).

\section{Pomeron loops rule.}
The question we have to ask is how do our numerical results of the previous section sit together with the picture of color electric fields having correlations on the scale of the saturation momentum, described in Sec. 2. Also, it is at first sight surprising that we do not find correlations in the target wave function which have been discussed analytically and numerically in \cite{yoshi} (albeit for a single dipole target).

To understand why this is the case, we note again that our calculation is done in the framework of the projectile dipole model, while that of \cite{yoshi} in the target dipole model. These two dipole models are large $N_c$ limits of the JIMWLK and KLWMIJ evolutions respectively, and in the rest of this section we will not distinguish between the corresponding evolution and its large $N_c$ limit. 

As discussed in \cite{foam} the JIMWLK and KLWMIJ evolutions are very different as far as the target wave function is concerned. KLWMIJ is the normal perturbative evolution of the target wave function. It is well known that in such evolution the number of gluons grows exponentially fast. The gluon density, which in this regime satisfies the BFKL equation, depends on the total rapidity as
\begin{equation}
g(p,Y)\propto e^{c\alpha_s Y}
\end{equation}
where $c$ is a number of order one and $Y$ is the total rapidity by which the target wave function has been boosted from "`rest". The exponential dependence is also true for the differential gluon density at any rapidity $\eta<Y$, namely (we suppress the momentum dependence in the prefactor for simplicity)
\begin{equation}
\frac{d}{d\eta}g(p,\eta)\propto e^{c\alpha_s \eta}\theta(Y-\eta)\,.
\end{equation}
Gluons in the wave function which are separated in rapidity by no more than $\delta\eta\sim O(\alpha_s)$ are correlated. Since the gluon density in the wave function exponentially grows with rapidity, any projectile that probes such a wave function effectively feels only the gluons in the last rapidity "`bin"'. Thus such a probe is sensitive to any correlation that exists between the softest gluons in the wave function.

Now consider JIMWLK evolution. The rapidity dependence of gluon density generated by the JIMWLK evolution is very different. As discussed in \cite{foam}, while the probability to emit an additional gluon in the dilute regime is proportional to the number of gluons in the wave function, in the dense regime this probability approaches a constant. The amplitude of the emission depends on the color fields in the target $E_i$, roughly as
\begin{equation}\label{amplitude1}
A\propto \frac{D_i}{D^2}E_i\,; \ \ \ \ \ \ \ \ \ \ \ \ \ \ \ \ \ \ \ D_i=\partial_i-gE_i
\end{equation}
In the dilute system this is proportional to the chromo-electric field $E$, while in the dense regime, where $D\sim E$ this is a constant. The evolution with constant probability of emission generates gluon density which is uniformly distributed in rapidity. This can be thought of as a random walk in color space as in \cite{foam}. So now one has 
\begin{equation}
\frac{d}{d\eta}g(p,\eta)= C\,.
\end{equation}
with $C$ a function of transverse momentum, but not of rapidity.
It is still true, like in the KLWMIJ case, that gluons separated by a small rapidity interval are correlated, while the correlation disappears for gluons at very different rapidities. Now, as opposed to KLWMIJ however if one scatters any projectile on such a target, the projectile will sample gluons at all rapidities equally. Thus, for example, if the projectile consists of two dipoles, the two dipoles will most likely scatter on color field components (gluons) with very different rapidities. Since such fields are not correlated, the two dipoles will scatter independently and the two dipole scattering amplitude will not exhibit any correlations. 

Another way of understanding the difference between the nature of JIMWLK and KLWMIJ evolutions is the following. In KLWMIJ evolution one starts initially with the target wave function which contains a small number of gluons. The probability to emit an additional gluon in one step of the evolution is small ($O(\alpha_s)$), however if a gluon is emitted, the end configuration strongly differs from the initial one, since the number of gluons have changed by a factor of order unity. Such evolution thus generates a very rough ensemble of target field (gluon) configurations. In this ensemble configurations with very different properties are present, albeit with small weight. Such an ensemble must exhibit large fluctuations in a variety of observables. 

On the other hand, in JIMWLK evolution the probability to emit an extra gluon is large - of order unity. However emission of an extra gluon produces a new configuration which is hardly distinct from the existing one, since the one extra gluon is produced on the background of $\frac{1}{\alpha_s}$ gluons which already exist in the wave function. Thus the JIMWLK ensemble is very different - it contains many configurations, but all these configurations have very similar properties. The fluctuations in observables in such an ensemble are very small.

Thus we conclude that KLWMIJ evolution must produce (and preserve) correlations (and fluctuations) between scattering amplitudes of two projectile dipoles. These are indeed the correlations discussed in \cite{yoshi}. On the other hand JIMWLK evolution must lead to disappearance of any correlations initially present in the target ensemble, as we have seen in our calculations.

Given that the two approximation to high energy evolution lead to such qualitatively different answers for the observable we are interested in, naturally one should ask which one of them, if any should be used in quantitative calculations. Naively one might think, that since we are interested in high multiplicity situation, the target is dense and JIMWLK evolution is more suitable. This however is not the case. Recall that we expect the correlations to arise due to scattering at close values of impact parameter. The transverse distances in question should be smaller or of the order of the saturation radius $R_s$. This means that the scattering occurs on the components of the target color field with transverse momenta $p<Q_s$. However at these momenta the target wave function by definition is still dilute.  Even for a dense target JIMWLK evolution is not appropriate for all wavelengths. Referring to eq.(\ref{amplitude1}) we see that the amplitude is independent of the field only when we can neglect the derivative relative to the field $gE$ in the expression for covariant derivative, in other words only for momenta greater than the saturation momentum. Thus in the dense target, we expect the gluon density to behave roughly as
\begin{equation}
\frac{d}{d\eta}g(p,\eta)\propto e^{c\alpha_s \eta}\theta(p-Q_s(Y))+C\  \theta(Q_s(Y)-p)\,.
\end{equation}
At large momenta $p>Q_s$ the appropriate evolution is KLWMIJ and not JIMWLK.

Thus to be able to describe fluctuations one needs to be able to evolve the high momentum modes according to KLWMIJ while low momentum modes according to JIMWLK. In fact, the situation is even more complicated. One expects that the largest contribution to correlations comes from the modes with $p$ of order $Q_s$, since in the relative distances of order $R_s$ should dominate the integral over impact parameter. For these modes however, the Pomeron merging and splitting (JIMWLK and KLWMIJ contributions) are of equal importance. In other words, this momentum range is sensitive at the leading order to the Pomeron loop contributions. We thus conclude that Pomeron loops need to be included for proper treatment of this question.

\section*{Acknowledgments}

We would like to thank Genya Levin, Larry McLerran, Guilherme Milhano, Amir Rezaeian, Edward Shuryak, Anna Stasto, and Raju Venugopalan for  inspiring discussions relevant to this work.
The work of AK is supported by DOE grant DE-FG02-92ER40716.
The work of ML is partially supported by the Marie Curie Grant  PIRG-GA-2009-256313 and UCONN's Guest Professorship Award.

%%%%%%%%%%%%%%%%%%%%%%%%%%%%%%%%%%%%%%%%%%%%%%%%%%%%%%%%%%%%%%%%%%%%%%%

\end{document}